\documentclass[aps,prl,twocolumn,floats,final,superscriptaddress,citeautoscript,longbibliography]{revtex4-2}
\usepackage{multirow}
\usepackage{hhline}
\usepackage{enumerate}
\usepackage{amssymb,bm}
\usepackage{graphicx}
\usepackage{amsmath}
\usepackage{braket}
\usepackage{mathtools}
\usepackage{bbm}
\usepackage{booktabs}
\usepackage{tabularx}
\usepackage[colorlinks=true,linkcolor=blue,citecolor=blue,urlcolor=blue]{hyperref}
\usepackage{xcolor,soul}
\usepackage{changes}
\usepackage{lineno,comment} 
\usepackage{enumitem}

\newcommand{\Tr}{\operatorname{Tr}}

\newcommand*\diff{\mathop{}\!\mathrm{d}}

\newcommand{\VF}{\Upsilon}
\newcommand{\tVF}{\widetilde{\Upsilon}}
\newcommand{\R}{{\mathrm{R}}}
\newcommand{\A}{{\mathrm{A}}}

\begin{document}
	
	\title{Error-resilient Reversal of Quantum Chaotic Dynamics Enabled by Scramblons}
	
	
	\author{Yu-Chen Li}
	\thanks{These authors contribute equally to this work}
	\affiliation{Laboratory of Spin Magnetic Resonance, School of Physical Sciences,
		Anhui Province Key Laboratory of Scientific Instrument Development and Application,
		University of Science and Technology of China, Hefei 230026, China}
	
	\author{Tian-Gang Zhou}
	\thanks{These authors contribute equally to this work}
	\affiliation{Institute for Advanced Study, Tsinghua University, Beijing 100084, China}
	
	\author{Shengyu Zhang}
	\thanks{These authors contribute equally to this work}
	\affiliation{Laboratory of Spin Magnetic Resonance, School of Physical Sciences,
		Anhui Province Key Laboratory of Scientific Instrument Development and Application,
		University of Science and Technology of China, Hefei 230026, China}
	\affiliation{Hefei National Laboratory, Hefei 230088, China}
	
	\author{Ze Wu}
	\affiliation{Department of Physics, The Chinese University of Hong Kong, Hong Kong, China}
	\affiliation{Laboratory of Spin Magnetic Resonance, School of Physical Sciences,
		Anhui Province Key Laboratory of Scientific Instrument Development and Application,
		University of Science and Technology of China, Hefei 230026, China}
	
	\author{Liqiang Zhao}
	\affiliation{Laboratory of Spin Magnetic Resonance, School of Physical Sciences,
		Anhui Province Key Laboratory of Scientific Instrument Development and Application,
		University of Science and Technology of China, Hefei 230026, China}
	\affiliation{Hefei National Laboratory, Hefei 230088, China}
	
	\author{Haochuan Yin}
	\affiliation{Laboratory of Spin Magnetic Resonance, School of Physical Sciences,
		Anhui Province Key Laboratory of Scientific Instrument Development and Application,
		University of Science and Technology of China, Hefei 230026, China}
	\affiliation{Hefei National Laboratory, Hefei 230088, China}
	
	\author{Xiaoxue An}
	\affiliation{Laboratory of Spin Magnetic Resonance, School of Physical Sciences,
		Anhui Province Key Laboratory of Scientific Instrument Development and Application,
		University of Science and Technology of China, Hefei 230026, China}
	\affiliation{Hefei National Research Center for Physical Sciences at the Microscale, Hefei 230026, China}
    
	\author{Hui Zhai}
	\email{hzhai@tsinghua.edu.cn}
	\affiliation{Institute for Advanced Study, Tsinghua University, Beijing 100084, China}
	\affiliation{Hefei National Laboratory, Hefei 230088, China}
	
	\author{Pengfei Zhang}
	\email{pengfeizhang.physics@gmail.com}
	\affiliation{Department of Physics, Fudan University, Shanghai 200438, China}
	\affiliation{State Key Laboratory of Surface Physics, Fudan University, Shanghai 200438, China}
	\affiliation{Hefei National Laboratory, Hefei 230088, China}
	
	\author{Xinhua Peng}
	\email{xhpeng@ustc.edu.cn}
	\affiliation{Laboratory of Spin Magnetic Resonance, School of Physical Sciences,
		Anhui Province Key Laboratory of Scientific Instrument Development and Application,
		University of Science and Technology of China, Hefei 230026, China}
    \affiliation{Hefei National Research Center for Physical Sciences at the Microscale, Hefei 230026, China}
	\affiliation{Hefei National Laboratory, Hefei 230088, China}

	\author{Jiangfeng Du}
	\affiliation{Hefei National Laboratory, Hefei 230088, China}
        \affiliation{State Key Laboratory of Ocean Sensing and School of Physics, Zhejiang University, Hangzhou 310058, China}

    \begin{abstract}
	The emergence of the arrow of time in quantum many-body systems stems from the inherent tendency of Hamiltonian evolution to scramble quantum information and increase entanglement. While, in principle, one might counteract this temporal directionality by engineering a perfectly inverted Hamiltonian to reverse entanglement growth, such a scenario is fundamentally unstable because even minor imperfections in the backward evolution can be exponentially amplified, a hallmark of quantum many-body chaos. Therefore, successfully reversing quantum many-body dynamics demands a deep understanding of the underlying structure of quantum information scrambling and chaotic dynamics. In this letter, by using solid-state nuclear magnetic resonance on a macroscopic ensemble of randomly interacting spins, we measure the out-of-time-ordered correlator (OTOC) and validate key predictions of scramblon theory, a universal theoretical framework for information scrambling. Crucially, this theory enables us to isolate and mitigate errors in the OTOC caused by imperfections in the backward evolution. As a result, this protocol uncovers the anticipated exponential behavior of quantum many-body chaos and extracts the quantum Lyapunov exponent in a many-body experimental system for the first time. Our results push the fundamental limits of dynamical reversibility of complex quantum systems, with implications for quantum simulation and metrology.
    \end{abstract}
    
	\date{\today}
  	\maketitle

	Quantum information scrambling lies at the heart of thermalization in isolated quantum systems~\cite{DeutschQuantumStatisticalMechanics1991,SrednickiApproachThermalEquilibrium1999,Rigol2007ThermalizationAI,Kaufman2016QuantumTT,Swingle:2018ekw}. When a low-entanglement initial state evolves under a Hamiltonian $\hat{H}$, local information spreads across the entire system, becoming irretrievable via local measurements. At the same time, the entanglement entropy grows, rendering the system locally a thermal state. Intriguingly, one can envision a gedanken experiment: if we apply the reversed Hamiltonian $-\hat{H}$ to this thermal state, is it possible to recover the original low-entanglement state and its local information? This question echoes a profound analogy in black hole physics: whether information thrown in and thermalized by a black hole can ever be retrieved. Now, advances in quantum control techniques on various synthetic quantum platforms attempt to turn such a thought experiment into a tangible ambition~\cite{KaiserLocalizationdelocalizationTransitionDynamics2015,DuMeasuringOutofTimeOrderCorrelators2017a,CappellaroExploringLocalizationNuclear2018a,CappellaroEmergentPrethermalizationSignatures2019,LiExperimentalObservationEquilibrium2020,sanchez2020perturbation,dominguez2021decoherence,PastawskiEmergentDecoherenceInduced2022,DuEmergentUniversalQuench2024a,ReyMeasuringOutoftimeorderCorrelations2017a,MonroeVerifiedQuantumInformation2019,ReyUnifyingScramblingThermalization2019b,LinkeExperimentalMeasurementOutofTimeOrdered2022,ChenInformationScramblingQuantum2021,DuanInformationScramblingDynamics2022,OliverProbingQuantumInformation2022,ZhaoProbingOperatorSpreading2022,ZobristConstructiveInterferenceEdge2025,VuleticTimereversalbasedQuantumMetrology2022,VuleticImprovingMetrologyQuantum2023,ChinQuantumSimulationUnruh2019,ThomasEnergyResolvedInformationScrambling2021,LiObservationQuantumInformation2024a,YouObservationAnomalousInformation2024a,WeidemullerTimereversalDipolarQuantum2024,gao2025signal}, raising an even deeper question: can engineered inverted dynamics defy the arrow of time?
	
\begin{figure*}
		\centering
		\includegraphics[width=0.95\linewidth]{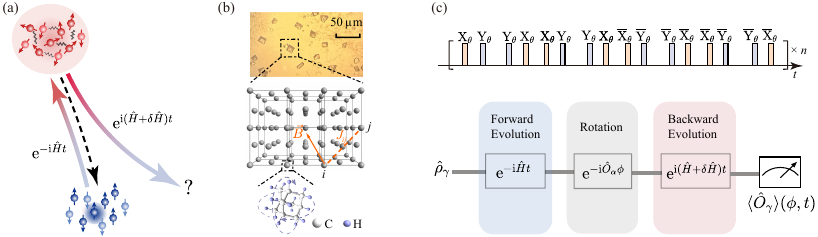}
		\caption{Schematic of Experimental Protocol.
			(a) Chaotic quantum many-body dynamics: Unitary evolution under $\hat{H}$ scrambles quantum information, transforming a low-entanglement initial state into a highly entangled state. Under time-reversed dynamics with perturbed Hamiltonian $-\hat{H}+\delta\hat{H}$, the system exhibits exponential deviation from perfect state recovery. 
			(b) Microscopic structure of adamantane (C$_{10}$H$_{16}$) powder: Each granule contains adamantane molecules arranged in a face-centered cubic lattice. Individual molecules consist of spin-$1/2$ $^1$H nuclei and spinless $^{12}$C atoms. 
			(c) \textit{Top}: Floquet pulse sequence converting the dipolar Hamiltonian Eq.~\eqref{dipolar} into the engineered form Eq.~\eqref{Heff} \cite{suter1987multiple,CappellaroExploringLocalizationNuclear2018a}. \textit{Bottom}: Experimental sequence for MQC and OTOC measurements. 
		}
		\label{fig:schematic}
	\end{figure*}

	In realistic quantum systems, however, such an oversimplified scenario does not materialize. As illustrated by Fig.~\ref{fig:schematic}(a), even infinitesimal imperfections in the engineered reversed Hamiltonian can lead to an exponentially amplified error as evolution time increases, a hallmark of quantum many-body chaos~\cite{shenkerBlackHolesButterfly2014,robertsLocalizedShocks2015,shenkerStringyEffectsScrambling2015,Maldacena_2016,kitaev2015simple}. Given that no practical implementation can perfectly realize $-\hat{H}$, the inherent chaotic nature of quantum many-body dynamics fundamentally prohibits the retrieval of information from a thermal state. The same reasoning extends analogously to the challenge of recovering information from a black hole. Recent experimental observations, spanning nuclear magnetic resonance (NMR) systems~\cite{KaiserLocalizationdelocalizationTransitionDynamics2015,DuMeasuringOutofTimeOrderCorrelators2017a,CappellaroExploringLocalizationNuclear2018a,CappellaroEmergentPrethermalizationSignatures2019,LiExperimentalObservationEquilibrium2020,sanchez2020perturbation,dominguez2021decoherence,PastawskiEmergentDecoherenceInduced2022,DuEmergentUniversalQuench2024a}, trapped ions~\cite{ReyMeasuringOutoftimeorderCorrelations2017a,MonroeVerifiedQuantumInformation2019,ReyUnifyingScramblingThermalization2019b,LinkeExperimentalMeasurementOutofTimeOrdered2022}, superconducting qubits~\cite{ChenInformationScramblingQuantum2021,DuanInformationScramblingDynamics2022,OliverProbingQuantumInformation2022,ZhaoProbingOperatorSpreading2022,ZobristConstructiveInterferenceEdge2025}, ultracold atoms~\cite{VuleticTimereversalbasedQuantumMetrology2022,VuleticImprovingMetrologyQuantum2023,ChinQuantumSimulationUnruh2019,ThomasEnergyResolvedInformationScrambling2021,LiObservationQuantumInformation2024a,YouObservationAnomalousInformation2024a,WeidemullerTimereversalDipolarQuantum2024}, and nitrogen vacancy centers~\cite{gao2025signal}, have also directly observed the inevitable effects of imperfect backward evolution. 
	
	Although errors in time-reversed dynamics are inherently unavoidable, a potential solution is to post-process experimental data to extrapolate an error-free limit. However, achieving an unbiased extrapolation remains highly challenging unless the imperfections meet the following two criteria: (1) the error model in the backward evolution must follow a universal behavior, independent of microscopic details; and (2) this universal behavior can be captured by a theoretical framework that yields an analytical ansatz. Such an ansatz would enable both the reliable quantification of imperfections in the backward evolution and robust extrapolation to the error-free limit.

	Remarkably, these two conditions can be simultaneously satisfied across a broad class of systems. Quantum information scrambling and many-body chaos are intimately linked, both characterized by the OTOC~\cite{larkin1969quasiclassical,shenkerBlackHolesButterfly2014,robertsLocalizedShocks2015,shenkerStringyEffectsScrambling2015,kitaev2015simple,Maldacena_2016}. The recently developed scramblon theory has identified the scramblon, a novel type of collective excitation in quantum many-body systems with all-to-all connectivity, as the fundamental carrier of scrambling dynamics \cite{SuhSoftModeSachdevYeKitaev2018,KitaevRelationMagnitudeExponent2019,yaoSubleadingWeingartens2022,ZhangTwowayApproachOutoftimeorder2022}. Analogous to phonons mediating density fluctuations or spin waves governing spin dynamics, scramblons dictate the spread of quantum information. The scramblon propagator grows exponentially with time, outpacing all other modes and dominating the late-time behavior of the OTOC, thereby resulting in fast information scrambling and a universal exponential behavior of the OTOC. Notably, in systems with holographic duality, scramblons in the boundary quantum theory correspond to gravitons in the bulk gravity \cite{SuhSoftModeSachdevYeKitaev2018,maldacenaConformalSymmetryIts2016}. 
	
	In this work, we measure the OTOC in a solid-state NMR sample with a macroscopic number of randomly interacting spins, a system beyond the capabilities of classical simulation. This can be achieved using a well-established protocol for measuring the multiple quantum coherence (MQC) spectrum (Fig.~\ref{fig:schematic}(c), bottom panel) \cite{yen1983multiple,baum1985multiple}. The measurement involves the reverse evolution of this many-body system, which, of course, is imperfect. We utilize scramblon theory to derive a fitting ansatz for the measured OTOC, achieving good agreement with the experiment. Moreover, the scramblon theory imposes several strong constraints on the fitting parameters, and we verify that these constraints are well satisfied by the fitting results. In other words, the experimental data strongly validate the scramblon theory. With this fitting formula, we can extrapolate the measured OTOC to the error-free limit. We demonstrate that the OTOC after this error mitigation exhibits the anticipated long-time behavior, in sharp contrast to other methods explored in the previous literature\cite{ChenInformationScramblingQuantum2021,ReyMeasuringOutoftimeorderCorrelations2017a,CappellaroEmergentPrethermalizationSignatures2019}. 
	
	\begin{figure*}
		\centering
		\includegraphics[width=0.95\linewidth]{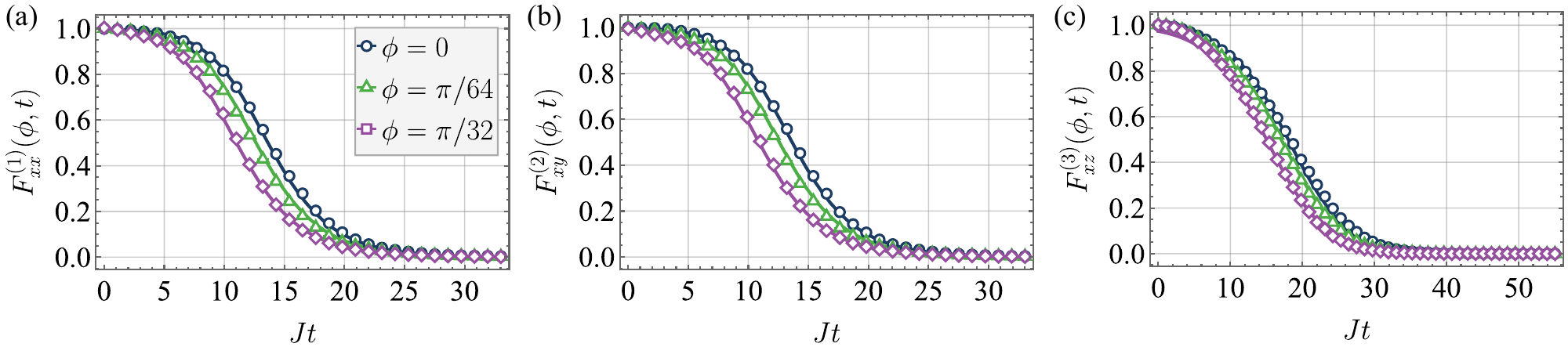}
		\caption{Fitting Experimental Data with the Scramblon Ansatz. Three typical experimental data $F^{(\lambda)}_{\alpha\gamma}(\phi,t)$ for three different $\phi=0$, $\pi/64$ and $\pi/32$. The superscript $\lambda$ labels three different Hamiltonians, as described in the main text. The solid curves are fitting results obtained using the scramblon ansatz. Experimentally, the $95\%$ confidence intervals determined from read-out noise are approximately $10^{-4}$, and therefore the error bars are contained within the data markers.}
		\label{fig:fitting}
	\end{figure*}
	
	\textit{Physical system and experimental protocol}---The experiment is performed on a powdered adamantane (C$_{10}$H$_{16}$) sample, as illustrated in Fig.~\ref{fig:schematic}(b) and detailed in the supplementary material~\cite{SM}. Each adamantane molecule contains 16 nuclear spin-$1/2$ hydrogen nuclei, denoted by $\mathbf{S}_{im}$, where the index $i$ labels the molecules and $m=1,\dots,16$ identifies individual spins within a molecule. Due to rapid molecular thermal motion, intramolecular spin interactions are averaged out, leaving only the intermolecular dipolar coupling as the dominant spin-spin interaction.
	The sample is placed in a uniform $9.4$ T magnetic field, which aligns the spin quantization axes and induces fast Larmor precession of all spins. In the rotating frame and under the secular approximation, the effective dipolar Hamiltonian takes the form \cite{duer2008solid}:
	\begin{equation}
		\hat{H}_\text{d}= \sum_{i<j,m,n} J_{ij} \left( -\hat{S}^x_{im}\hat{S}^x_{jn} - \hat{S}^y_{im}\hat{S}^y_{jn} + 2\hat{S}^z_{im}\hat{S}^z_{jn} \right).
		\label{dipolar}
	\end{equation}
	Here, the coupling strengths $J_{ij}$ are treated as random variables with zero mean ($\overline{J_{ij}}=0$) and variance $\overline{J^2_{ij}}=4J^2/N$, where $N$ is the total number of spins and $J \sim 2\pi \times 1460$ Hz \cite{DuEmergentUniversalQuench2024a}. This randomness of $J_{ij}$ arises from the anisotropic nature of dipolar interaction and the random crystalline orientations in the powder sample~\cite{SM}.

	Then, by applying periodic global radio-frequency pulse sequences to rotating spins, as depicted in Fig.~\ref{fig:schematic}(c), a more general effective Hamiltonian can be engineered via the Floquet-Magnus expansion as $\hat{H}_\text{eff}=\hat{H}_0+\hat{H}_1$. $\hat{H}_0$ represents the leading-order contribution in the expansion, which can be generally written as:
	\begin{equation}
		\hat{H}_{0} = \sum_{i<j,m,n}\sum_{\mu,\nu}J_{ij}\xi_{\mu\nu}\hat{S}^\mu_{im}\hat{S}^\nu_{jn}, \label{Heff}
	\end{equation}
	where $\mu,\nu = x,y,z$. The protocol generating Eq.~\eqref{Heff} imposes specific constraints on the coefficients $\xi_{\mu\nu}$, namely the traceless condition $\sum_{\mu=x,y,z}\xi_{\mu\mu}=0$ and the symmetry condition $\xi_{\mu\nu}=\xi_{\nu\mu}$. An important observation is that if a set $\{\xi_{\mu\nu}\}$ satisfies these constraints, so does the set $\{-\xi_{\mu\nu}\}$. This implies that Floquet engineering always enables reversal of $\hat{H}_0$. However, the higher-order terms, denoted by $\hat{H}_1$,  generally cannot be simultaneously reversed. Therefore, without loss of generality, we denote the forward evolution as governed by $\hat{H}=\hat{H}_0+\hat{H}_1$, and the backward evolution as governed by $-\hat{H}_0 - \hat{H}_1^\prime=-\hat{H}-\delta\hat{H}$. Here, $\delta\hat{H}$, representing the difference between $\hat{H}_1$ and $\hat{H}^\prime_1$, is the imperfection in the reversed dynamics. In this system, $\hat{H}_1$ {and $\delta\hat{H}$ are} typically on the order of a few percent relative to $\hat{H}_0$ {\cite{SM,DuEmergentUniversalQuench2024a}}.
	
	The measurement protocol is schematically illustrated in Fig.~\ref{fig:schematic}(c). We begin by defining the total spin operator along the $\gamma$-axis as $\hat{O}_\gamma=\sum_{im}\hat{S}^\gamma_{im}$, where $\gamma \in \{x, y, z\}$. The system is initialized in a weakly polarized state described by the density matrix $\hat{\rho}_\gamma \propto \mathbbm{1} + \epsilon\hat{O}_\gamma$, with $\epsilon \ll 1$ characterizing the small polarization. Following forward time evolution under the Hamiltonian $\hat{H}$, the density matrix transforms as $\hat{\rho}_\gamma(t) = {\rm e}^{-{\rm i}\hat{H}t}\hat{\rho}_\gamma {\rm e}^{{\rm i}\hat{H}t}$.
	Subsequently, we implement a uniform spin rotation by an angle $\phi$ around the $\alpha$-axis ($\alpha \in \{x, y, z\}$), yielding the transformed density matrix:
	$\hat{\rho}_{\alpha\gamma}(\phi,t) = {\rm e}^{-{\rm i}\hat{O}_\alpha\phi}{\rm e}^{-{\rm i}\hat{H}t}\hat{\rho}_\gamma {\rm e}^{{\rm i}\hat{H}t}{\rm e}^{{\rm i}\hat{O}_\alpha\phi}$.
	This is followed by backward evolution as described above. Finally, we measure the total magnetization along the $\gamma$-direction $\langle\hat{O}_\gamma\rangle$, which equals the correlation function $\epsilon F_{\alpha \gamma}(\phi,t)$ defined as:
	\begin{equation}
		F_{\alpha \gamma}(\phi,t)=\text{Tr}[\hat{O}_\gamma(0)\hat{\mathcal{V}}^\dagger(t) {\rm e}^{-{\rm i}\phi\hat{O}_\alpha (t)}\hat{O}_\gamma(0){\rm e}^{{\rm i}\phi\hat{O}_\alpha (t)}\hat{\mathcal{V}}(t)]. \label{Fphit}
	\end{equation}
	Here, $\hat{\mathcal{V}}(t) \equiv \mathbf{T}\exp\left(-{\rm i}\int_0^t{\rm d} t'\delta\hat{H}(t')\right)$ represents the time-evolution operator in the interaction picture, with $\mathbf{T}$ denoting time ordering. All operators $\hat{O}(t)$ and $\delta\hat{H}(t)$ evolve according to the Heisenberg picture under $\hat{H}$. For normalization, we divide $F_{\alpha \gamma}(\phi,t)$ by $\mathcal{C} = \langle \hat{O}_\gamma^2\rangle = N/4$ to ensure $F_{\alpha \gamma}(0,0) = 1$. The MQC spectrum can be obtained through the Fourier transformation of $F_{\alpha \gamma}(\phi,t)$ with respect to $\phi$ \cite{yen1983multiple,baum1985multiple}.
	
	\textit{Validating the scramblon theory}---We now highlight the relationship between $F_{\alpha \gamma}(\phi,t)$ and OTOCs through two limits: (i) For perfect reversed dynamics ($\delta\hat{H}=0$), $F_{\alpha\gamma}(\phi,t)$ reduces to
	\begin{equation}\label{reduction1}
		F_{\alpha \gamma}^{(0)}(\phi,t)=\text{Tr}[ \hat{O}_\gamma(0) {\rm e}^{-{\rm i}\phi\hat{O}_\alpha (t)}\hat{O}_\gamma(0){\rm e}^{{\rm i}\phi\hat{O}_\alpha (t)}],
	\end{equation}
	representing an OTOC of operators ${\rm e}^{{\rm i}\phi\hat{O}_\alpha(t)}$ and $\hat{O}_\gamma$; (ii) At $\phi=0$, it becomes
	\begin{equation}\label{reduction2}
		F_{\alpha \gamma}(0,t)=\text{Tr}[\hat{O}_\gamma(0)\hat{\mathcal{V}}^\dagger(t)\hat{O}_\gamma(0)\hat{\mathcal{V}}(t)],
	\end{equation}
	and this OTOC, also known as the Loschmidt echo, probes imperfections in the reversed dynamics. Both cases can be computed by scramblon theory \cite{ZhangTwowayApproachOutoftimeorder2022,ZhangSignatureScramblonEffective2024}, strongly suggesting that the general behavior of $F_{\alpha \gamma}(\phi,t)$, although more challenging to derive, can also be captured by scramblon theory. Indeed, as we show in the Supplementary Material~\cite{SM}, an ansatz for $F_{\alpha \gamma}(\phi,t)$ in the small-$\phi$ regime can be derived as:
	\begin{equation}
		F_{\alpha \gamma}(\phi,t)= \frac{1}{\left(1+a {\rm e}^{\varkappa t}+b \phi^2 {\rm e}^{\varkappa t}\right)^{2\Delta}},\label{fitting}
	\end{equation}
	where $\{a,b,\varkappa,\Delta\}$ are all fitting parameters that depend on
$\alpha$ and $\gamma$. The $\phi=0$ limit of Eq.~\eqref{fitting} reproduces the Loschmidt echo (Eq.~\eqref{reduction2}), with $a\neq 0$ quantifying the reversal imperfection. We note that a recent theoretical work also discussed a similar issue in a specific solvable model \cite{swingleScramblingDynamicsImperfections2025}.
	
	\begin{figure*}
		\centering
		\includegraphics[width=0.95\linewidth]{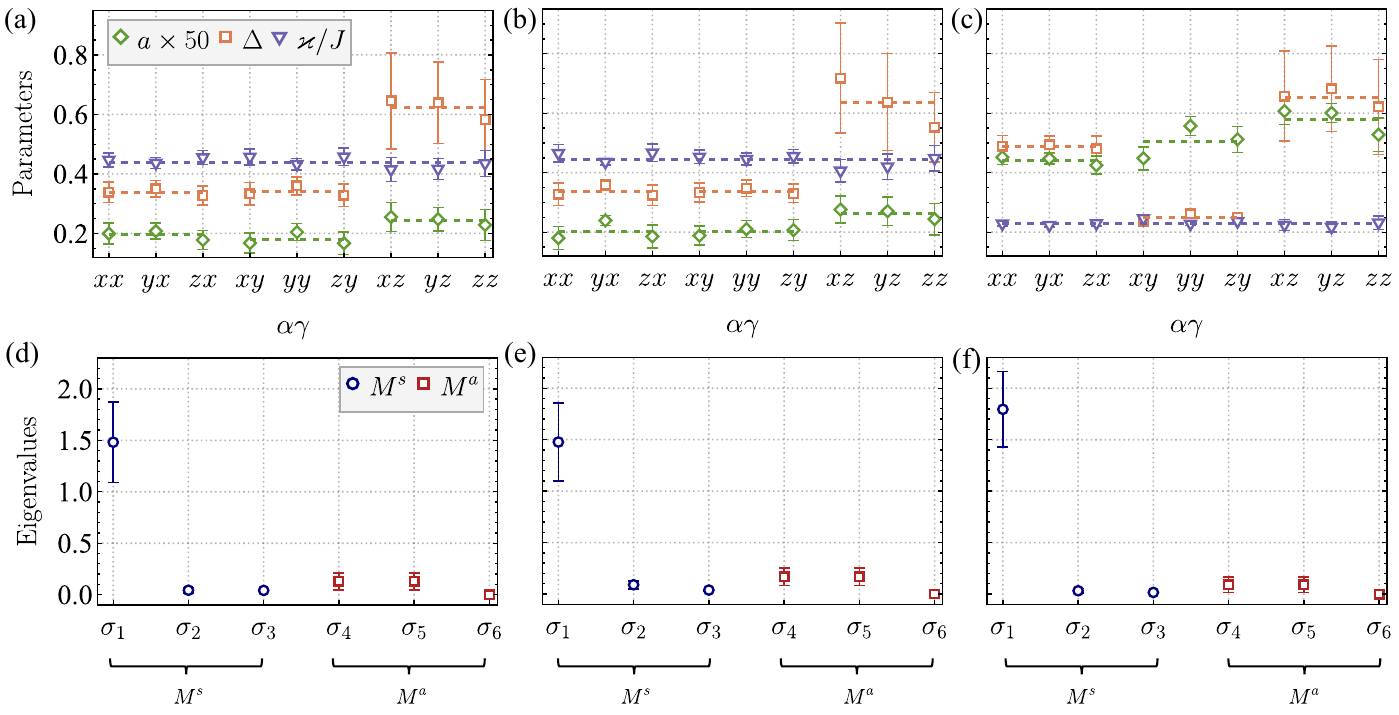}
		\caption{Verifying the Predictions of Scramblon Theory. The three columns correspond to the fitting parameters extracted from three different Hamiltonians: $\hat{H}^{(1)}$, $\hat{H}^{(2)}$, and $\hat{H}^{(3)}$. Subfigures (a)-(c) in the upper row display the fitting parameters for different values of $\alpha$ and $\gamma$, with green diamonds representing $a\times50$, yellow squares representing $\Delta$, and blue triangles representing $\varkappa$. The dashed lines represent $\varkappa$ averaged over $\alpha$ and $\gamma$, and $a$ and $\Delta$ averaged over $\alpha$. Subfigures (d)-(f) in the lower row show the eigenvalues $\sigma_1,\sigma_2,\sigma_3$ for the symmetric matrix $M^\text{s}$ and the absolute value of eigenvalues $\sigma_4,\sigma_5,\sigma_6$ for the antisymmetric matrix $M^\text{a}$. The error bars on the fitting parameters represent the $95\%$ confidence interval, including the standard deviation from each fitting residual $\sigma_{\text{res}}$ and the uncertainty arising from the choice of fitting range $\sigma_{\text{range}}$~\cite{SM}.}
		\label{fig:verification}
	\end{figure*}
	
	We investigate three representative chaotic Hamiltonians: (i) $\hat{H}_0^{(1)}$ with $\xi_{xx}=-\xi_{yy}=0.4$, (ii) $\hat{H}_0^{(2)}$ with $\xi_{xy}=\xi_{yx}=0.4$, and (iii) $\hat{H}_0^{(3)}$ with $(\xi_{xx},\xi_{yy},\xi_{zz})={(0.2,-0.225,0.025)}$, while all other $\xi_{\mu\nu}=0$. For each Hamiltonian $\hat{H}^{(\lambda)}_0$, we perform measurements for all nine possible combinations of $\alpha,\gamma\in\{x,y,z\}$ axes, yielding a total of $3\times9=27$ distinct cases. For each case, we measure three different $\phi$ angles and fit the three curves jointly with a single scramblon ansatz, Eq.~\eqref{fitting}.
	We find that all data are excellently captured by the scramblon ansatz Eq.~\eqref{fitting}, with three representative fits shown in Fig.~\ref{fig:fitting}. We quantify the goodness of fit using the adjusted $R^2$, as detailed in the Supplementary Material~\cite{SM}. Discrepancies between experiment and theory may arise from the invalidity of the scramblon theory at very early times $e^{\varkappa t}\lesssim 1$.
	
	The scramblon theory not only provides the fitting ansatz but also imposes strong constraints on the parameters. For a given Hamiltonian, while the fitting yields parameter sets indexed by $\alpha$ and $\gamma$, the theory predicts the following:
	
	\begin{enumerate}[label=\arabic*.]
		
		\item $\varkappa$ must be independent of both $\alpha$ and $\gamma$, since the system is dominated by a single scramblon mode and $\varkappa$ characterizes the intrinsic growth rate of this mode.
		
		\item $a$ and $\Delta$ cannot depend on $\alpha$ as they are retained in the $\phi=0$ limit where no $\alpha$-axis spin rotation is applied.
		
		\item The matrix $M_{\alpha\gamma} \equiv b_{\alpha\gamma}\Delta_\gamma$ must be symmetric and rank-1, reflecting the symmetric coupling of a single scramblon mode to both $\hat{O}_\alpha$ and $\hat{O}_\gamma$ operators.
		
	\end{enumerate}
	
	Fig.~\ref{fig:verification} demonstrates how we verify these three constraints for our three Hamiltonians. Panels (a-c) display the values of $a$, $\Delta$, and $\varkappa$ for all nine measurement combinations per Hamiltonian, with dashed lines indicating (i) $\varkappa$ averaged over $\gamma$ and $\alpha$, and (ii) $\alpha$-averaged $a$ and $\Delta$ values that depend only on $\gamma$. Most cases show only a few percent deviations from averaged values, with a maximum deviation of {$\sim 16\%$} occasionally, confirming the first two constraints. To test the third constraint, we decompose $M$ into symmetric ($M^\text{s}=(M+M^\text{T})/2$) and antisymmetric ($M^\text{a}=(M-M^\text{T})/2$) components. Eigenvalues shown in panels (d-f) reveal that $M^\text{s}$ exhibits only one dominant eigenvalue, and all eigenvalues of $M^\text{a}$ are negligible, validating the rank-1 symmetric requirement for $M$. It is notable that a strong correspondence between the results for $H^{(1)}_0$ and $H^{(2)}_0$ is observed. This is anticipated, as $H^{(1)}_0$ can be transformed into $H^{(2)}_0$ via a global rotation ${\rm e}^{-{\rm i}\hat{O}_z\pi/4}$.
	
	\begin{figure}
		\centering
		\includegraphics[width=0.95\linewidth]{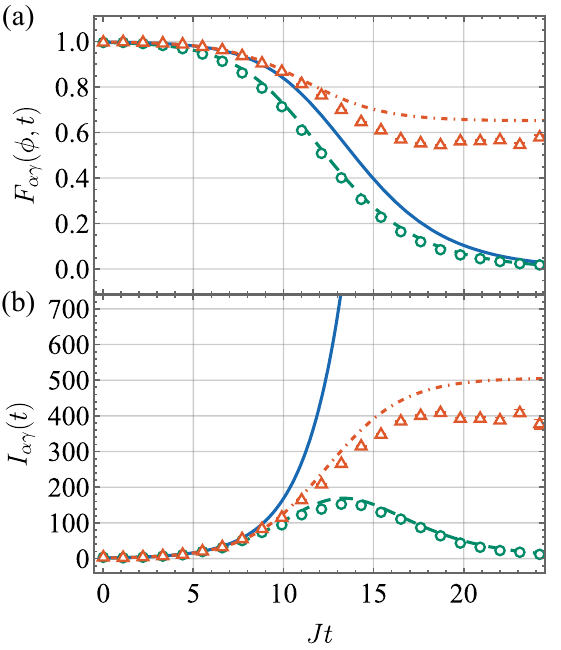}
		\caption{Error Mitigation of the Reversed Quantum Many-Body Dynamics. (a) $F_{\alpha\gamma}(\phi, t)$ and (b) $I_{\alpha\gamma}(t)$ defined in Eq.~\eqref{eq:OTOC_commutator}. As an example, we show $\alpha,\gamma=z,y$, $\phi=\pi/64$ and with Hamiltonian $\hat{H}^{(1)}_0$. The blue solid line results from the scramblon ansatz by setting $a=0$. The red triangles are error mitigated by $\tilde{F}_{\alpha\gamma}(\phi,t)$, and green circles are raw data without error mitigation. The red dash-dotted line and the green dashed line are obtained from the scramblon ansatze of $\tilde{F}_{\alpha\gamma}(\phi,t)$ and $F_{\alpha\gamma}(\phi,t)$ (without setting $a=0$). The experimental data points in (b) were extracted by fitting the corresponding data for $F_{\alpha\gamma}(\phi,t)$ with a sixth-order polynomial. The errorbars represent the standard deviations.}
		\label{fig:mitigation}
	\end{figure}
	
	\textit{Recovering exponential behavior of OTOC and extracting the Lyapunov exponent}---The excellent agreement with experimental data strongly validates the scramblon ansatz. Within this framework, the parameter $a$ fully captures imperfections in the time-reversed dynamics, as evidenced by $F_{\alpha\gamma}(0,t)\equiv 1$ when $a=0$. This allows an error mitigation by setting $a=0$ in Eq.~\eqref{fitting} while keeping other fitted parameters unchanged (blue curve in Fig.~\ref{fig:mitigation}(a)). For comparison, we examine an alternative mitigation approach from previous literatures that normalizes $F_{\alpha\gamma}(\phi,t)$ by $F(0,t)$ to define $\tilde{F}_{\alpha\gamma}(\phi,t)=F_{\alpha\gamma}(\phi,t)/F_{\alpha\gamma}(0,t)$ (red triangles in Fig.~\ref{fig:mitigation}(a)){~\cite{ChenInformationScramblingQuantum2021,ReyMeasuringOutoftimeorderCorrelations2017a,CappellaroEmergentPrethermalizationSignatures2019,YaoOperatorGrowthOpen2023}}. Within the scramblon formalism, this yields:
	\begin{equation}
		\tilde{F}_{\alpha \gamma}(\phi,t)= \left(\frac{1+a {\rm e}^{\varkappa t}}{1+a {\rm e}^{\varkappa t}+b \phi^2 {\rm e}^{\varkappa t}}\right)^{2\Delta}.\label{fitting2}
	\end{equation}
	Different approaches are compared in Fig.~\ref{fig:mitigation}(a).
	
	To assess the reliability of different approaches, we compute $I_{\alpha\gamma}(\phi)$ from $F_{\alpha\gamma}(\phi,t)$ as:
	\begin{equation}
		I_{\alpha\gamma}(t) \equiv -\partial_{\phi}^2 F_{\alpha\gamma}(\phi,t)|_{\phi=0}, \label{eq:OTOC_commutator}
	\end{equation}
	which for perfect time reversal ($\delta\hat{H}=0$) corresponds to the OTO commutator $\langle|[\hat{O}_\alpha(t),\hat{O}_\gamma]|^2\rangle$ and exhibits exponential growth, reflecting the rapid increase of the operator size $\hat{O}_\alpha(t)$ under Heisenberg evolution before reaching the scrambling time. Analysis based on the scramblon ansatz reveals: (i) The raw data (Eq.~\eqref{fitting}) gives $I_{\alpha\gamma}(t)=4b\Delta {\rm e}^{\varkappa t}/(1+a {\rm e}^{\varkappa t})^{1+2\Delta}$, which decays at late times due to $\Delta>0$ (green dashed line in Fig.~\ref{fig:mitigation}(b)); (ii) The normalized approach (Eq.~\eqref{fitting2}) yields $I_{\alpha\gamma}(t)=4b\Delta {\rm e}^{\varkappa t}/(1+a {\rm e}^{\varkappa t})$, saturating to $4b\Delta/a$ at long times (red dash-dotted line in Fig.~\ref{fig:mitigation}(b)); and (iii) Only our $a=0$ mitigation produces the physically expected exponential growth $4b\Delta {\rm e}^{\varkappa t}$ (blue solid line in Fig.~\ref{fig:mitigation}(b)), where $\varkappa$ emerges as the quantum Lyapunov exponent. It is also remarkable that only our $a=0$ mitigated result $I_{\alpha\gamma}(t)$ exhibits the symmetric and rank-1 property predicted by the scramblon theory for systems with time-reversal symmetry, because there is a single dominant scramblon mode and its coupling to both $\hat{O}_\alpha$ and $\hat{O}_\gamma$ operators is symmetric (see Supplementary Materiral \cite{SM} for more experimental data on all combinations of $\alpha$ and $\gamma$). These distinct behaviors are quantitatively compared with experimental data.  To the best of our knowledge, $\varkappa$ shown in Fig.~\ref{fig:verification} serves as the first exact experimental determination of the Lyapunov exponent in a quantum many-body system with macroscopic degrees of freedom. 
	
	\textit{Conclusion}---In conclusion, we develop an ansatz based on scramblon theory to describe reversed quantum many-body dynamics in the presence of imperfections. We validate this ansatz by fitting it to experimental data from an NMR system with macroscopically randomly interacting spins, where the stringent constraints imposed by scramblon theory are convincingly verified by the fitting outputs. These results suggest that the applicability of the scramblon theory extends beyond idealized all-to-all interacting models, remaining valid in realistic systems with microscopic long-range interactions. The application of the scramblon theory allows us to mitigate the effects of imperfections and recover error-resilient reversed dynamics. Notably, time-reversal-based exponential sensitivity has also been recently exploited to enhance quantum metrology in systems such as cold atoms in cavities~\cite{VuleticTimereversalbasedQuantumMetrology2022,VuleticImprovingMetrologyQuantum2023} and nitrogen vacancy centers~\cite{gao2025signal}, where imperfections also suppress this sensitivity at long evolution times. Our theory may be extended to such quantum metrology applications, potentially enabling further improvements in sensitivity.
	
	\ 
	
	\ 
	
	\textit{Acknowledgements}---This work is supported by Quantum Science and Technology-National Science and Technology Major Project 2021ZD0302005 (H.Z.), 2024ZD0300101 (P.Z.), and  2021ZD0303205 (X.P.), National Natural Science Foundation of China No.~12488301 (H.Z.), No.~U23A6004 (H.Z.), No.~12374477 (P.Z.), No.~12261160569 (X.P.), the Shanghai Rising-Star Program under grant number 24QA2700300 (P.Z.), and the XPLORER Prize (H.Z. and X.P.). We thank experimental assistance from the Instruments Center for Physical Science, University of Science and Technology of China.
	\vspace{1em}
		
	\textit{Data Availability}---The data that support the findings of this study are available at Zenodo (https://doi.org/10.5281/zenodo.16142455)~\cite{code_zenodo}. Further data are available from the corresponding authors upon reasonable request. 


%

	\newpage

	\section{End matter}
	\textit{Hamiltonian Engineering}---An effective Hamiltonian $\hat{H}_{\text{eff}}$ can be engineered from the dipolar Hamiltonian $\hat{H}_{\text{d}}$ (Eq.~\eqref{dipolar}), through applying the radio-frequency (RF) pulse sequences periodically. The RF pulse train was constructed from a basic four-pulse sequence applied from left to right in chronological order \cite{suter1987multiple,CappellaroExploringLocalizationNuclear2018a}:
	\begin{equation}
		\left(\tau_1, \mathbf{X}_\theta, \tau_2, \mathbf{Y}_\theta, \tau_3, \mathbf{Y}_\theta, \tau_4, \mathbf{X}_\theta, \tau_5\right),
		\label{sequence}
	\end{equation}
	where $\mathbf{X}_\theta=\mathrm{e}^{-\mathrm{i}\hat{O}_z\theta}\mathrm{e}^{-\mathrm{i}\hat{O}_x\pi/2}\mathrm{e}^{\mathrm{i}\hat{O}_z\theta}$ denotes a $\pi/2$ pulse with azimuthal phase $\theta$; the operator $\mathbf{Y}_\theta$ is defined similarly for a rotation about the $y$-axis. The terms $\{\tau_g\}$ represent the inter-pulse delays. 
	
	The average Hamiltonian can be calculated in a toggling frame following the discrete rotations induced by the pulses \cite{duer2008solid}. In this frame, the Hamiltonian becomes a piecewise-constant function in time intervals. During a given interval $\tau_g$, the toggling-frame Hamiltonian is:
	\begin{equation}
		\hat{H}_{\text{togg}}^{\{g\}} =\sum_{i<j,m,n} J_{ij}[3(\hat{\bm{S}}_{im} \cdot \bm{u}_g) (\hat{\bm{S}}_{jn} \cdot \bm{u}_g) - \hat{\bm{S}}_{im}\cdot \hat{\bm{S}}_{jn}],
	\end{equation}
	where the unit vector $\bm{n}_g$ is determined by the preceding pulses. We designed the pulse intervals $\tau_g$ and phases $\theta$ such that the zeroth-order average Hamiltonian matched the target Hamiltonian $\hat{H}_{0}$ (Eq.~\eqref{Heff}), which is expressed as $\hat{H}_{0} = \frac{1}{T}\sum_{g=1}^5\tau_g\hat{H}_{\text{togg}}^{\{g\}}$, where $T=\sum_{g=1}^5\tau_g$ is the total duration of the pulse sequence. The resulting anisotropic tensor $\bm{\xi}$ has elements given by $\xi_{\mu\nu} =\sum_{g=1}^{5} \xi^{\{g\}}_{\mu\nu} \tau_g/T$, where $\xi_{\mu\nu}^{\{g\}} = 3 u_g^{\mu}u_g^{\nu} - \delta_{\mu\nu}$ is the term during the interval $\tau_g$.
	
	The elements of the time-averaged tensor $\bm{\xi}$ are subject to the following constraints. The diagonal terms are traceless and bounded:
	\begin{equation}
		\sum_{\mu} \xi_{\mu\mu} = 0, \quad -1\le\xi_{\mu\mu}\le2\ \ \ \text{(Constraint 1)}.
	\end{equation}
	The off-diagonal terms are symmetric and are also bounded:
	\begin{equation}
		\xi_{\mu\nu} = \xi_{\nu\mu},\quad |\xi_{\mu\nu}| \leq 3/2 \ \ \ \ \text{(Constraint 2, for }\mu\neq\nu\text{)}.
	\end{equation}
	Experimental limitations, specifically a finite pulse width ($2~\mu\text{s}$ in our case), prevent the delays $\tau_g$ from reaching zero, which limits the accessible range of the tensor elements. For a cycle period of $T = 30~\mu\text{s}$, the achievable ranges were restricted to $-0.6\le\xi_{\mu\mu}\le1.2$ and $|\xi_{\mu\nu}| \leq 0.9$ ($\mu\ne\nu$). By tuning $\{\tau_g\}$ and $\theta$, we successfully engineered all target $\hat{H}_0$ with both positive and negative signs. Further details are provided in the Supplementary Material \cite{SM}.
	
	The forward evolution is captured by the total effective Hamiltonian $\hat{H}=\hat{H}_0+\hat{H}_1$, while the one for the backward evolution is $-\hat{H}_0-\hat{H}_1'$. The deviation from ideal time-reversal is $\delta\hat{H}=\hat{H}_1'-\hat{H}_1$, arising from the higher-order Floquet-Magnus expansions { and experimental control imperfections such as finite pulse width, flip-angle errors and field fluctuations}. We quantify the magnitude of $\delta\hat{H}$ with {an error metric $\varepsilon\equiv\sqrt{\Tr[(\delta\hat{H})^2]/\Tr[(\hat{H}_0)^2]}$.} Numerical simulations up to the second order of Floquet-Magnus expansion and with $N=8$ spins suggests that {$\varepsilon \sim 3\%$} for the three different $\hat{H}_0$ investigated (see Supplementary Material \cite{SM} for details). 
	
	\textit{Scramblon Field Theory}---Scramblon field theory provides an analytical framework for deriving various generalizations of OTOC~\cite{SuhSoftModeSachdevYeKitaev2018,KitaevRelationMagnitudeExponent2019,yaoSubleadingWeingartens2022,ZhangTwowayApproachOutoftimeorder2022}. For example, for a general OTOC involving simple operators $\hat{W}$ and $\hat{V}$, it reads:
 	\begin{equation}
 		\label{eq:OTOC1}
 		\begin{aligned}
 			F_{\hat{W},\hat{V}}
 			&=\langle \hat{W}^\dagger(t_1) \hat{V}^\dagger(t_3)\hat{W}(t_2) \hat{V}(t_4) \rangle \\ &=
 			\includegraphics[scale=0.9]{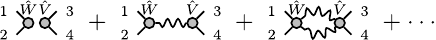}
 			\\
 			&=\sum_{k=0}^{\infty}\frac{(-\lambda)^{k}}{k!}\VF^{\R,k}_{\hat{W}}(t_1-t_2)\VF^{\A,k}_{\hat{V}}(t_3-t_4).
 		\end{aligned}
 	\end{equation}
The vertices $\VF^{\R,k}_{\hat{W}}$ and $\VF^{\A,k}_{\hat{V}}$, represented by the gray dots, are the scattering vertices between operators $\hat{W}$, $\hat{V}$, and $k$ scramblons in the future and the past, respectively. To ensure that the scramblon enters the free collective mode region, we set $t_1\approx t_2 \gg t_3\approx t_4$. The term $\lambda= C^{-1} e^{\varkappa \frac{t_1+t_2-t_3-t_4}{2}}$ is the propagator for the scrambling modes, depicted by the wavy lines. The prefactor $C$ is proportional to the number of degrees of freedom $N$.

 	As discussed in the main text, two types of OTOCs from Eq.~\eqref{reduction1} and Eq.~\eqref{reduction2} are involved in the calculation of the correlator $F_{\alpha\gamma}(\phi,t)$.
This implies that both operators $\hat{O}_\alpha$ and $\delta\hat{H}$ can exchange scramblons with the operator $\hat{O}_{\gamma}$.
 		Leaving further details in the supplementary material~\cite{SM}, we show that these contributions can be captured by the equation:
\begin{equation}\label{eq:F_simplify_result}
 		\begin{split}
 			F_{\alpha \gamma}(\phi, t)
 			&=f^{\A}_{\hat{O}_{\gamma}} \Big(\tilde{\VF}^{\R,1}_{\delta \hat{H}}C^{-1} \varkappa^{-1} e^{\varkappa t} \\
 			&\qquad\ \ + \phi^2 \VF^{\R,1}_{\hat{O}_{\alpha}}(0) C^{-1}e^{\varkappa t}, 0\Big) \mathcal{C}^{-1} . \\
 		\end{split}
 	\end{equation}
 	This is the central prediction of the scramblon theory. The constant $\tVF^{\R,1}_{\delta\hat{H}} = \int_{-\infty}^{\infty} \diff{t'} \VF^{\R,1}_{\delta\hat{H}}(t')$ is the reduced vertex function and $\mathcal{C}$ is the normalization constant. Here we have introduced the auxiliary function $ f^{\R/\A}(x,t) \equiv \sum_{k=0}^\infty \frac{(-x)^k}{k!} \VF^{\R/\A,k}(t)$ to simplify the result. In the argument of Eq.~\eqref{eq:F_simplify_result}, the first term corresponds to the contribution from imperfection, while the second term arises from the uniform spin rotation.
 	
 	In general, obtaining an analytical form for $f^{\R}_{\hat{O}_{\gamma}}(x,t)$ in randomly interacting spin systems is difficult. However, all known solvable models, such as the large-$q$ SYK, Brownian SYK, and Brownian circuit models, share a similar ansatz for $f^{\R}_{\hat{O}_{\gamma}}(x,t)$ as~\cite{ZhangTwowayApproachOutoftimeorder2022}:
 \begin{equation}\label{eq:f_ansatz}
 		f^R_{\hat{O}_{\gamma}}(x,0) = f^A_{\hat{O}_{\gamma}}(x,0) = \mathcal{C} \left(1 + c_{\gamma} x\right)^{-2\Delta_{\gamma}},
 	\end{equation}
 	where the parameters $c_{\gamma}$ and the scaling dimension $\Delta_{\gamma}$ depend on both the choice of model and the detection direction $\gamma$. Under this assumption, Eq.\eqref{eq:F_simplify_result} reduces to the form of Eq.\eqref{fitting}, with the following identification:
 \begin{equation}\label{eq:parameters}
 		a_{\gamma} = c_{\gamma} C^{-1}\varkappa^{-1}\tilde{\VF}^{\R,1}_{\delta \hat{H}}, \ \ \ b_{\alpha\gamma} = \VF^{\R,1}_{\hat{O}_{\alpha}}(0) c_{\gamma} C^{-1}.
 	\end{equation}
 	Furthermore, the vertex function of the spin operators is related to the ansatz in Eq.~\eqref{eq:f_ansatz} by the relation $\VF^{\R/\A,1}_{\hat{O}_{\gamma}}(t) = -\partial_x f^{\R/\A}_{\hat{O}_{\gamma}}(x,t)|_{x=0}$, yielding $\VF^{\R/\A,1}_{\hat{O}_{\gamma}} = 2\Delta_{\gamma}c_{\gamma}$. We can then immediately verify all the constraints introduced in the main text. In particular, we find that $M_{\alpha\gamma}=b_{\alpha \gamma} \Delta_\gamma = 2 \Delta_{\alpha}c_{\alpha}\Delta_{\gamma}c_{\gamma}\mathcal{C}C^{-1}$ is a symmetric matrix in $\alpha$ and $\gamma$, which must satisfy $\text{rank}(M_{\alpha\gamma})=1$.

\end{document}


\title{Supplementary Information}


\maketitle
\tableofcontents
\clearpage
\section{Experimental System}

\subsection{System Hamiltonian}
The experiments were conducted on a 400-MHz NMR spectrometer. The sample is powdered adamantane (C\textsubscript{10}H\textsubscript{16}), and our investigation focuses on the network of spin-1/2 $^{1}$H nuclei. The most abundant carbon isotope, $^{12}$C (99\% natural abundance), is spin-zero and does not contribute to the NMR signal. The sample was maintained at room temperature within a homogeneous magnetic field of $B_0=9.4$~T, aligned along the $z$-axis and generated by a superconducting magnet. The spin-lattice relaxation time was measured to be $T_1 = 1.3$~s.

At room temperature, adamantane exists in a face-centered cubic (fcc) plastic crystal phase (space group Fm3m)~\cite{nordman1965phase,mccall1960nuclear,chang1960heat}. In this phase, the nearly spherical molecules undergo rapid, isotropic reorientation on a picosecond timescale ($\sim 10^{-11}$~s)~\cite{resing1969nmr}. This rapid tumbling averages the intra-molecular dipole-dipole interactions to zero. Consequently, the nuclear spin dynamics are dominated by the inter-molecular dipolar couplings between protons on different molecules, which can be roughly estimated via approximating the spin locations by their respective molecular centers~\cite{resing1969nmr,mccall1960nuclear,smith1961calculation}.

In the laboratory frame, the total Hamiltonian for the $^{1}$H spin system is:
\begin{equation}
    \hat{H}_{\text{lab}}=\hat{H}_{\rm Z}+\hat{H}_{\mathrm{CS}}+\hat{H}_{\rm rf}(t)+\hat{H}_{\mathrm{dip}}.
    \label{LabSysHam}
\end{equation}
The terms represent the Zeeman interaction ($\hat{H}_{\rm Z}$), chemical shifts and field inhomogeneities ($\hat{H}_{\mathrm{CS}}$), the time-dependent radio-frequency pulses ($\hat{H}_{\rm rf}(t)$), and the dipolar interaction ($\hat{H}_{\mathrm{dip}}$).
The first term, $\hat{H}_{\rm Z} = \omega_{\rm H}\hat{O}_z$, is the dominant Zeeman interaction, where the Larmor frequency is $|\omega_{\rm H}|/(2\pi) =|\gamma_{\rm H}B_0|/(2\pi) = 400.15$~MHz ($\gamma_{\rm H}$ is the proton gyromagnetic ratio) and $\hat{O}_z = \sum_{im} \hat{S}_{im}^z$, with $\hat{S}_{im}^\alpha$ ($\alpha=x,y,z$) the spin operators for the protons ($m=1,2,3,\cdots,16$) in molecule $i$. The second term, $\hat{H}_{\mathrm{CS}}$, represents small frequency offsets due to chemical shifts and magnetic field inhomogeneities. The chemical shift distribution for adamantane is on the order of tens of Hertz~\cite{SDBSWeb}, and field inhomogeneities contribute a line broadening of less than 12~Hz. These small terms are effectively refocused by the multi-pulse sequences used and are neglected hereafter.

The third term represents the spin Hamiltonian induced by the radio-frequency (RF) pulse, which reads
\begin{equation}
    \hat{H}_{\rm rf}(t)=\omega_{1}[\cos(\omega_{\rm rf}t+\theta)\hat{O}_{x}+\sin(\omega_{\rm rf}t+\theta)\hat{O}_{y}].
\end{equation}
Here $\omega_1$ is the nutation frequency, and $\theta$ is the pulse phase. We operate at the resonant condition $\omega_{\rm rf}=\omega_{\rm H}$. Finally, the forth term, $\hat{H}_{\mathrm{dip}}$, is the nuclear dipole-dipole interaction:
\begin{equation}
\hat{H}_{\mathrm{dip}}=\sum\limits_{i<j,m,n}\frac{\mu_0\hbar^2\gamma^2_\text{H}}{4\pi R^3_{ij}}\Bigg[\hat{\bm{S}}_{im}\cdot\hat{\bm{S}}_{jn}-\frac{3(\hat{\bm{S}}_{im}\cdot{\bf R}_{ij})(\hat{\bm{S}}_{jn}\cdot{\bf R}_{ij})}{R^2_{ij}}\Bigg].
\end{equation}
Here, $\mu_0$ is the vacuum magnetic permeability and $\gamma_\text{H}$ is the proton's gyromagnetic ratio. The vector $\mathbf{R}_{ij}$ connects the centers of molecules $i$ and $j$ with distance $R_{ij} = |\mathbf{R}_{ij}|$. 

In the limit where $|\omega_{\rm H}| \gg \omega_1$ and $|\omega_{\rm H}| \gg \|\hat{H}_{\rm dip}\|$, it is conventional to move to the interaction picture with respect to $\hat{H}_{\rm Z}$ (equivalent to a rotating frame analysis) and retains only the components of $\hat{H}_{\mathrm{dip}}$ that commute with $\hat{O}_z$. For the radio-frequency pulse, the dominant contribution is 
\begin{equation}
    \hat{H}_{\rm rf}^{\rm rot}=\omega_{1}[\hat{O}_{x}\cos\theta+\hat{O}_{y}\sin\theta].
\end{equation}
For the dipole-dipole interaction, this yields the secular dipolar Hamiltonian~\cite{abragam1961principles,duer2008solid}, which governs the system's evolution during the delays between RF pulses:
\begin{align}
    \hat{H}_{\mathrm{d}} \equiv \sum_{i<j,m,n} J_{ij} (3\hat{S}^{z}_{im}\hat{S}^{z}_{jn} - \hat{\bm{S}}_{im}\cdot\hat{\bm{S}}_{jn}),
    \label{sysHam}
\end{align}
where the effective coupling is
\begin{equation}
    J_{ij} = \left(\frac{\mu_0}{4\pi}\right)\frac{\hbar\gamma_{\mathrm{H}}^{2}}{2R_{ij}^3}\left(1-3\cos^2\theta_{ij}\right).
    \label{eq:def_Jij}
\end{equation}
$\theta_{ij}$ is the angle between $\mathbf{R}_{ij}$ and the external magnetic field direction $z$. Putting all the ingredients together, we can express the total system Hamiltonian in the rotating frame as
\begin{equation}
     \hat{H}=\hat{H}_{\rm rf}^{\rm rot}+\hat{H}_{\mathrm{d}}.
\end{equation}  
In the experiment, we apply discrete pulses of finite duration $\tau_{\rm p} = 2\ \mu\text{s}$, satisfying $\omega_1 \tau_{\rm p} = \pi/2$, to induce $\pi/2$ rotations along directions specified by the phase $\theta$. 
During the intervals between pulses, the radio-frequency field is turned off, yielding $\omega_1 = 0$.

\subsection{Characterization of the Dipolar Coupling $J_{ij}$}
\subsubsection{Numerical Estimation from Lattice Structure}
The coupling $J_{ij}$ can be treated as a random variable in a powdered sample because the crystallographic axes are randomly oriented with respect to the external magnetic field. To characterize its distribution, we numerically calculated $J_{ij}$ values based on the fcc lattice structure of adamantane (lattice constant $a_0=0.945$~nm~\cite{nordman1965phase}). 

In our model, the locations of a lattice site $j$ relative to a central site $i$ are given by $\mathbf{R}_{ij}=n_1\mathbf{a}_1+n_2\mathbf{a}_2+n_3\mathbf{a}_3$, where $\{\mathbf{a}_k\}$ are the principal axes of the lattice ($|a_1|=|a_2|=|a_3|=a_0$). For a given orientation of the external magnetic field $\mathbf{B}_0$ relative to these axes, we calculated $J_{ij}$ using Eq.~\eqref{eq:def_Jij}. The distances between the center cite $i$ and the serial nearest neighbors are $a_0\sqrt{2}/2,\ a_0,\ a_0\sqrt{6}/2,\cdots$. 
We included couplings up to the 13th nearest neighbors, a distance of $a_0\sqrt{26}/2\approx 2.4$~nm, which encompasses $N_{\rm m}=320$ lattice sites. At this distance, the coupling strength decays to about $2\%$ of the nearest-neighbor value on average, validating that interactions are short-range compared to the powder granule size ($\sim \mu$m).
Therefore, we can assume that the calculation of $J_{ij}$ is identical for different choices of $i$.

\begin{figure}[htb]
    \centering
    \includegraphics[width=0.5\linewidth]{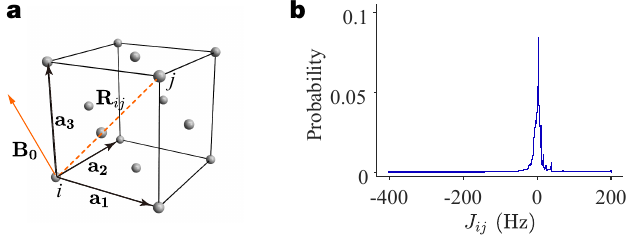}
    \caption{\textbf{Probability distribution of the intermolecular coupling $J_{ij}$.} \textbf{a}, Schematic of the calculation. Adamantane molecules are modeled as point-like dipole sources on an fcc lattice. The coupling $J_{ij}$ between a pair $(i,j)$ depends on their connecting vector $\mathbf{R}_{ij}$ and its orientation relative to the external field $\mathbf{B}_0$. \textbf{b}, The calculated distribution of $J_{ij}$ values, summarized over $10^5$ random crystal orientations and including interactions up to the 13th nearest-neighbor shell.}
    \label{fig:Jij_distribution}
\end{figure}

The resulting probability distribution of $J_{ij}$, averaged over $10^5$ random crystal orientations, is plotted in \ref{fig:Jij_distribution}. Two characteristics should be noted. Firstly, the powder average guarantees that $\overline{J_{ij}}=0$ since 
\begin{equation}
    \overline{1-3\cos^2\theta}=\int_0^{\pi}\dd \theta(1-3\cos^2\theta)\sin\theta =0.
\end{equation}
Secondly, the distribution is asymmetric, a consequence of the range of the angular term, $(1-3\cos^2\theta_{ij}) \in [-2, 1]$. We define the effective dipolar coupling strength $J$ as
\begin{equation}
    J\equiv\sqrt{\frac{N\overline{J_{ij}^2}}{4}}=2\sqrt{N_{\rm m}\overline{J_{ij}^2}}=2\sqrt{N_{\rm m}\frac{\sum_{j,j\ne i}^{N_{\rm m}-1}\overline{J_{ij}^2}}{N_{\rm m}-1}}\approx 2\sqrt{\sum_{j,j\ne i}^{N_{\rm m}-1}\overline{J_{ij}^2}}=(\frac{\mu_0}{4\pi})\hbar\gamma_{\rm H}^2\sqrt{\sum_{j,j\ne i}^{N_{\rm m}-1}\frac{\overline{(1-3\cos\theta_{ij}^2)^2}}{R_{ij}^6}}.
    \label{eq:J_def}
\end{equation}
Here $N=16N_{\rm m}$ is the total number of spins.
The integral
\begin{equation}
    \overline{(1-3\cos\theta_{ij}^2)^2}=\frac{1}{2}\int_0^\pi\dd\theta_{ij}(1-3\cos\theta_{ij}^2)^2\sin\theta_{ij}=\frac{4}{5},
\end{equation}
so
\begin{equation}
J=\frac{\mu_0}{2 \pi} \frac{\hbar \gamma_{\mathrm{H}}^2}{\sqrt{5}} \sqrt{\sum_{j, j \neq i}^{N_{\mathrm{m}}-1} \frac{1}{R_{i j}^6}}.
\end{equation}
For face-centered cubic lattice, the summation below is convergent~\cite{jones1925calculation,kittel2018introduction}:
\begin{equation}
    \lim_{N_{\rm m}\to\infty}\sqrt{\sum_{j, j \neq i}^{N_{\rm m}-1} \frac{1}{R_{i j}^6}}\approx\sqrt{\frac{115.63136}{a_0^6}}.
\end{equation}
Therefore, the numerically estimated effective dipolar coupling strength $J_{\rm nume}\approx1369$ Hz.

\subsubsection{Experimental Characterization of $J$}
The effective dipolar coupling strength, $J$, was also determined experimentally by analyzing the free induction decay (FID) of the $^{1}$H spins~\cite{DuEmergentUniversalQuench2024a}. The FID is a normalized autocorrelation function of the transverse magnetization such as 
\begin{equation}
    f(t)\equiv\frac{\Tr(\mathrm{e}^{-\mathrm{i}\hat{H}_{\mathrm{d}}t}\hat{O}_x\mathrm{e}^{\mathrm{i}\hat{H}_{\mathrm{d}}t}\hat{O}_x)}{\Tr(\hat{O}_x^2)}.
    \label{eq:Def_FID}
\end{equation}
The second-order derivative of FID
\begin{equation}
    M_2 \equiv -\left(\frac{\mathrm{d}^{2}f(t)}{\mathrm{d}t^{2}}\right)_{t=0} = -\frac{\Tr\left([\hat{H}_{\mathrm{d}},~\hat{O}_x]^2\right)}{\Tr(\hat{O}_x^2)},
\end{equation}
is directly related to the effective coupling strength $J$ via
\begin{equation}
    M_2 =\frac{9}{4}(N-1)\overline{J_{ij}^2}\approx9J^2.
    \label{eq:m2}
\end{equation}
This result can be derived following the formulas in Refs.~\cite{van1948dipolar,abragam1961principles}.
Therefore, by fitting the experimental FID and extracting its second-order derivative, one can determine $J$. The FID of adamantane is well-described by the function proposed by Abragam~\cite{abragam1961principles}:
\begin{equation}
    f_{\text{fit}}(t) = \exp\left(-\frac{a^2t^2}{2}\right)\frac{\sin(bt)}{bt}.
    \label{FID_fit}
\end{equation}
Fitting the FID to Eq.~\eqref{FID_fit} allows for the extraction of the second moment via $M_2 = a^2+b^2/3$. Ref.~\cite{DuEmergentUniversalQuench2024a} yields an experimental value of $J_{\rm exp}=\sqrt{M_2}/3 \approx 2\pi\times 1460$~Hz. The $\sim 7\%$ discrepancy between $J_{\rm exp}$ and $J_{\rm nume}$ could be attributed to the point-like molecule approximation in our calculation, which slightly underestimates the true coupling strength. We use the experimentally characterized dipolar coupling strength $J_{\rm exp}$ for the notation of the dimensionless time scale $Jt$.

\subsection{Signal and Uncertainty}
\begin{figure}[htb]
    \centering
    \includegraphics[width=0.3\linewidth]{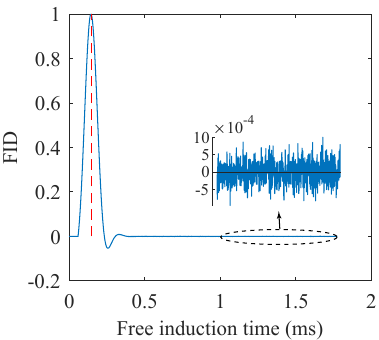}
    \caption{\textbf{FID curve.} A representative FID corresponding to $F_{xx}(0,0)$ acquired using the magic echo method. The value of $F_{xx}(0,0)$ is determined from the normalized amplitude at the echo peak (indicated by the red dashed line). The uncertainty of $F_{xx}(0,0)$ is calculated from the standard deviation of the final 100 data points in the signal tail.}
    \label{fig:FID}
\end{figure}

The collective magnetization of the final state $\hat{\rho}_{\rm f}$, $\langle\hat{O}_\gamma\rangle=\Tr(\rho_{\rm f}\hat{O}_\gamma)$ were determined from their corresponding FIDs. Before sampling the FID, a waiting time is necessary in order to get rid of the distortion from the excitation circuit. To compensate the signal decay due to evolution under $\hat{H}_{\rm d}$ during the waiting time, a magic echo pulse sequence was employed. This technique generates a time-reversed dipolar Hamiltonian, $-\hat{H}_{\rm d}$, which refocuses the magnetization to form an echo~\cite{rhim1970violation}. Additionally, a triple-pulse excitation scheme with phase cycling was implemented to suppress radio-frequency (RF) acoustic ringing artifacts~\cite{wang2021triple}. A representative FID acquired using these methods is shown in \ref{fig:FID}. 

For each measurement, the value of $\langle\hat{O}_\gamma\rangle$ was taken as the normalized amplitude at the peak of the magic echo, indicated by the dashed red line in \ref{fig:FID}. The uncertainty in this value was calculated as the standard deviation of the final 100 data points in the FID signal tail, a measure that includes the readout noise from the probe. The error bars presented in the figures were then determined through standard linear propagation of this uncertainty.

\subsection{State Initialization and Global Rotations}

In the absence of RF irradiation, the system resides in a thermal equilibrium state determined by the static magnetic field at room temperature:
\begin{equation}
    \hat{\rho}_0 = \frac{\mathrm{e}^{-\beta\hbar\hat{H}_{\rm lab}}}{\mathcal{Z}} \approx \frac{\mathrm{e}^{-\beta\hbar\omega_{\rm H}\hat{O}_z}}{\mathcal{Z}} \approx \frac{\mathbbm{1}-\beta\hbar\omega_{\rm H} \hat{O}_z}{2^N}.
    \label{eq:rho0}
\end{equation}
Here, $\beta=1/(k_{\rm B}T)$ is the inverse temperature and $\mathcal{Z}$ is the partition function. The first approximation holds because the Zeeman energy is much larger than the internal dipolar energies. The second step is the high-temperature approximation, valid since the dimensionless parameter $\epsilon = \beta\hbar\omega_{\rm H} \ll 1$ (at room temperature and 9.4~T, $\epsilon \sim 10^{-5}$). For convenience, we often write the initial state as $\hat{\rho}_0 \propto \mathbbm{1}+\epsilon\hat{O}_z$. A subsequent $\pi/2$ pulse along the $x$ or $y$ axis turns $\hat{\rho}_0$ into $\hat{\rho}_\gamma$ with $\gamma=y$ or $x$.  

\begin{figure}[htb]
    \centering
    \includegraphics[width=0.4\linewidth]{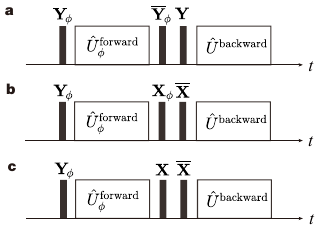}
    \caption{\textbf{Pulse sequences for implementing global rotations.} The diagrams show the additional pulses and phase shifts $\phi$ required to realize the rotation $\mathrm{e}^{-\mathrm{i}\hat{O}_\alpha\phi}$ for (\textbf{a}) $\alpha=x$, (\textbf{b}) $\alpha=y$, and (\textbf{c}) $\alpha=z$.}
    \label{fig:rotation}
\end{figure}

A key technique in our protocol is the ability to apply a global rotation $\mathrm{e}^{-\mathrm{i}\hat{O}_\alpha\phi}$ to the state after a period of forward evolution. This is achieved using the standard phase increment technique, which involves inserting a pair of $\pi/2$ pulses and applying a phase shift $\phi$ to specific pulses in the overall sequence~\cite{cho2003spin,ramanathan2003encoding}. This method fulfills a precise control on the rotation angle $\phi$. For example, the rotation $\mathrm{e}^{-\mathrm{i}\hat{O}_x\phi}$ is realized using the sequence in \ref{fig:rotation}a. The essential mechanism relies on the operator identity $\mathrm{e}^{-\mathrm{i}\hat{O}_x\phi} = \mathrm{e}^{-\mathrm{i}\hat{O}_y\pi/2}\mathrm{e}^{-\mathrm{i}\hat{O}_z\phi}\mathrm{e}^{\mathrm{i}\hat{O}_y\pi/2}$. By strategically applying the phase shifts, this composite rotation is constructed and applied to the state after it has evolved under $\hat{U}_{\rm forward}$. The total unitary operator is 
\begin{equation}
\begin{aligned}
    \hat{U}&=\hat{U}_{\rm backward}\mathrm{e}^{-\mathrm{i}\hat{O}_y\pi/2}\left(\mathrm{e}^{-\mathrm{i}\hat{O}_z\phi}\mathrm{e}^{\mathrm{i}\hat{O}_y\pi/2}\mathrm{e}^{\mathrm{i}\hat{O}_z\phi}\right)\left(\mathrm{e}^{-\mathrm{i}\hat{O}_z\phi}\hat{U}_{\rm forward}\mathrm{e}^{\mathrm{i}\hat{O}_z\phi}\right)\left(\mathrm{e}^{-\mathrm{i}\hat{O}_z\phi}\mathrm{e}^{-\mathrm{i}\hat{O}_y\pi/2}\mathrm{e}^{\mathrm{i}\hat{O}_z\phi}\right)\\
    &=\underbrace{\hat{U}_{\rm backward}}_{\text{Backward evolution}}  \underbrace{\mathrm{e}^{-\mathrm{i}\hat{O}_y\pi/2}\mathrm{e}^{-\mathrm{i}\hat{O}_z\phi}\mathrm{e}^{\mathrm{i}\hat{O}_y\pi/2}}_{\text{Composite rotation}} \underbrace{\hat{U}_{\rm forward}}_{\text{Forward evolution}} \underbrace{\mathrm{e}^{-\mathrm{i}\hat{O}_y\pi/2}}_{\text{Initial pulse}} \mathrm{e}^{\mathrm{i}\hat{O}_z\phi} \\
    &=\hat{U}_{\rm backward}\mathrm{e}^{-\mathrm{i}\hat{O}_x\phi}\hat{U}_{\rm forward}\mathrm{e}^{-\mathrm{i}\hat{O}_y\pi/2} \mathrm{e}^{\mathrm{i}\hat{O}_z\phi}.
\end{aligned}
\end{equation}
The first $\mathrm{e}^{\mathrm{i}\hat{O}_z\phi}$ operator can be omitted in the further time evolution since it has no effect on $\hat{\rho}_0$.
Thus, the initial pulse creates $\hat{\rho}_x\propto \mathbbm{1}+\epsilon\hat{O}_x$, which evolves under $\hat{U}_{\rm forward}$ and is then subjected to the desired global rotation $\mathrm{e}^{-\mathrm{i}\hat{O}_x\phi}$. The rotations $\mathrm{e}^{-\mathrm{i}\hat{O}_y\phi}$ and $\mathrm{e}^{-\mathrm{i}\hat{O}_z\phi}$ are realized using analogous pulse sequences, as depicted in \ref{fig:rotation}b and \ref{fig:rotation}c, respectively. The $\hat{O}_z$ rotation sequence includes an identity operation ($\mathbf{X}\overline{\mathbf{X}}$) to ensure all rotation schemes have the same duration, thus experiencing comparable levels of experimental imperfection.

\section{Hamiltonian Engineering}

\subsection{Pulse Sequence Design}
To engineer the target Hamiltonians, we periodically apply RF pulse sequences to the dipolar Hamiltonian Eq.~\eqref{sysHam}. The sequence we use is based on Suter8 sequence~\cite{suter1987multiple}:
\begin{equation}
	\Big(\tau_1, \mathbf{X}_\theta, \tau_2, \mathbf{Y}_\theta, \tau_3, \mathbf{Y}_\theta, \tau_2, \mathbf{X}_\theta, \tau_1\Big)\Big(\tau_1, \mathbf{X}_\theta, \tau_2, \mathbf{Y}_\theta, \tau_3, \mathbf{Y}_\theta, \tau_2, \mathbf{X}_\theta, \tau_1\Big),
    \label{eq:seq1}
\end{equation}
where $\mathbf{X}_\theta=\mathrm{e}^{-\mathrm{i}(\hat{O}_x\cos\theta+\hat{O}_y\sin\theta)\pi/2}=\mathrm{e}^{-\mathrm{i}\hat{O}_z\theta}\mathrm{e}^{-\mathrm{i}\hat{O}_x\pi/2}\mathrm{e}^{\mathrm{i}\hat{O}_z\theta}$ denotes a $\pi/2$ pulse (with width $\tau_{\text{p}}=2\ \mu\text{s}$) applied about an axis in the $xy$-plane with an azimuthal phase $\theta$. The operator $\mathbf{Y}_\theta$ is defined analogously with respect to the $y$-axis. The terms $\{\tau_g\}$ are the inter-pulse delays. The pulses are applied from left to right in chronological order. To enhance robustness against flip-angle errors, the Suter8 sequence is symmetrized by concatenating its copy $\overline{\rm Suter8}$ with conjugate pulse phases, $\mathbf{X}_\theta\to\overline{\mathbf{X}}_\theta=(\mathbf{X}_\theta)^\dagger=\mathbf{X}_{\theta+\pi}$, etc, to constitute the Wei16 sequence~\cite{CappellaroExploringLocalizationNuclear2018a}.

To clarify how to obtain the average Hamiltonian, we take the basic building block of Suter8 and Wei16, which is a four-pulse sequence:
\begin{equation}
	\left(\tau_1, \mathbf{X}_\theta, \tau_2, \mathbf{Y}_\theta, \tau_3, \mathbf{Y}_\theta, \tau_4, \mathbf{X}_\theta, \tau_5\right),
    \label{eq:seq1}
\end{equation}
with $\tau_1=\tau_5$ and $\tau_2=\tau_4$. Since the total spin operator $\hat{O}_z$ commutes with $\hat{H}_{\text{d}}$, the overall phase shift can be factored out, making the sequence equivalent to:
\begin{equation}
    \mathrm{e}^{-\mathrm{i}\hat{O}_z\theta}\left(\tau_1, \mathbf{X}, \tau_2, \mathbf{Y}, \tau_3, \mathbf{Y}, \tau_4, \mathbf{X}, \tau_5\right)\mathrm{e}^{\mathrm{i}\hat{O}_z\theta}.
    \label{eq:seq2}
\end{equation}
The average Hamiltonian is calculated within a toggling frame that rotates in synchrony with the applied pulses \cite{duer2008solid}. In this frame, the dipolar Hamiltonian transforms into a piecewise-constant operator:
\begin{equation}
    \hat{H}_{\text{togg}}(t)=\hat{U}_{\text{rf}}^\dagger(t)\hat{H}_{\text{d}}\hat{U}_{\text{rf}}(t),
\end{equation}
where $\hat{U}_{\text{rf}}(t)={\rm e}^{-{\rm i}\mathbf{T}\int_0^t\hat{H}_{\rm rf}^{\rm rot}\dd t}$ is the unitary propagator for the sequence of RF pulses applied up to time $t$. The Hamiltonian during the $g$-th interval, $\tau_g$, is given by:
\begin{equation}
	\hat{H}_{\text{togg}}^{\{g\}} =\sum_{i<j,m,n} J_{ij}[3(\hat{\bm{S}}_{im} \cdot \bm{u}_g) (\hat{\bm{S}}_{jn} \cdot \bm{u}_g) - \hat{\bm{S}}_{im}\cdot \hat{\bm{S}}_{jn}],
\end{equation}
where $\bm{u}_g=(u_g^x,u_g^y,u_g^z)$ is a unit vector representing the orientation of the spin quantization axis set by the preceding pulses. Within this framework, the effective Hamiltonian is derived using the Floquet-Magnus expansion, i.e., $\hat{H}=\hat{H}_{\text{0th}}+\hat{H}_{\text{1st}}+\hat{H}_{\text{2nd}}+\cdots$. We calculate up to the second order:
\begin{equation}
    \begin{aligned}
        \hat{H}_{\text{0th}} &\equiv \frac{1}{T}\sum_{g_1=1}^5\hat{H}_{\text{togg}}^{\{g_1\}}\tau_{g_1},\\
        \hat{H}_{\text{1st}} &\equiv \frac{-\mathrm{i}}{2 T}\sum_{g_1=1}^5\sum_{g_2=1}^{g_1}\left[\hat{H}_{\text{togg}}^{\{g_1\}}, \hat{H}_{\text{togg}}^{\{g_2\}}\right]\tau_{g_1}\tau_{g_2},\\
        \hat{H}_{\text{2nd}} &\equiv -\frac{1}{6 T} \sum_{g_1=1}^5\sum_{g_2=1}^{g_1}\sum_{g_3=1}^{g_2}\left\{\left[\hat{H}_{\text{togg}}^{\{g_1\}},[\hat{H}_{\text{togg}}^{\{g_2\}}, \hat{H}_{\text{togg}}^{\{g_3\}}]\right]+\left[\hat{H}_{\text{togg}}^{\{g_3\}},[\hat{H}_{\text{togg}}^{\{g_2\}}, \hat{H}_{\text{togg}}^{\{g_1\}}]\right]\right\}\tau_{g_1}\tau_{g_2}\tau_{g_3},
    \end{aligned}
    \label{eq:FM_expansion}
\end{equation}
where $T=\sum_{g=1}^5\tau_g$ is the total period of the pulse sequence. The norm of the $n$-th order term is on the order of $\mathcal{O}[(JT)^n]$. If $\hat{H}_{\text{togg}}(t)$ during one cycle is symmetric in time (i.e., a palindrome), the odd-order terms of the Floquet-Magnus expansion (Eq.~\eqref{eq:FM_expansion}) vanish due to the properties of nested commutators. In our case, the total effective Hamiltonian $\hat{H}=\hat{H}_{\text{0th}}+\hat{H}_{\text{2nd}}+\hat{H}_{\text{4nd}}+\cdots$.

\subsection{Parameters for Target Hamiltonians}
The experimental parameters $(\tau_g, \theta)$ are chosen such that the zeroth-order average Hamiltonian, $\hat{H}_{\text{0th}}$, matches the desired target Hamiltonian, $\hat{H}_{0}$:
\begin{equation}\label{eqn:H}
	\hat{H}_{0} \equiv \sum_{i<j,m,n} \sum_{\mu,\nu} J_{ij}\xi_{\mu\nu}\hat{S}_{im}^\mu \hat{S}_{jn}^\nu,
\end{equation}
where $\mu,\nu\in\{x,y,z\}$. The elements of the dimensionless anisotropy tensor are given by $\xi_{\mu\nu} =\sum_{g=1}^{5} \xi^{\{g\}}_{\mu\nu} \tau_g/T$, with 
\begin{equation}
    \xi^{\{g\}}_{\mu\nu}=3u_g^\mu u_g^\nu-\delta_{\mu\nu}.
\end{equation}
For simplicity, we first consider the case with no phase shift ($\theta=0$). The direction vectors $\bm{u}_g$ and the corresponding diagonal elements of the anisotropy tensor ${\bm{\xi}}^{\{g\}}$ for each interval are listed in \ref{tab:piecewise_Ham}.
\begin{table}[htb]
    \centering
    \begin{tabular}{c|c|c|c}
        \hline
        Interval & $\bm{u}_g$ & $(\xi_{xx}^{\{g\}}, \xi_{yy}^{\{g\}}, \xi_{zz}^{\{g\}})$ & $\xi_{\mu\nu}^{\{g\}}\ (\mu\ne\nu)$\\ 
        \hline
        $\tau_1$ & $(0,0,1)$ & $(-1,-1,2)$ & 0\\
        $\tau_2$ & $(0,1,0)$ & $(-1,2,-1)$ & 0\\
        $\tau_3$ & $(1,0,0)$ & $(2,-1,-1)$ & 0\\
        $\tau_4$ & $(0,1,0)$ & $(-1,2,-1)$ & 0\\
        $\tau_5$ & $(0,0,1)$ & $(-1,-1,2)$ & 0\\
        \hline
    \end{tabular}
    \caption{Piecewise Hamiltonian parameters during the pulse intervals of the sequence in Eq.~\eqref{eq:seq2} for $\theta=0$.}
    \label{tab:piecewise_Ham}
\end{table}
Averaging over the cycle period $T$ yields the total anisotropy tensor $\bm{\xi}$, which in this case is diagonal:
\begin{equation} \label{eq:xi_diag}
    \bm{\xi} = 
    \begin{bmatrix}
    \xi_{xx} & \xi_{xy} & \xi_{xz}\\
    \xi_{yx} & \xi_{yy} & \xi_{yz}\\
    \xi_{zx} & \xi_{zy} & \xi_{zz}\\
    \end{bmatrix}
    =
    \frac{1}{T}
    \begin{bmatrix}
    2\tau_3 - (\tau_1+\tau_2+\tau_4+\tau_5) & 0 & 0\\
    0 & 2(\tau_2+\tau_4) - (\tau_1+\tau_3+\tau_5) & 0\\
    0 & 0 & 2(\tau_1+\tau_5) - (\tau_2+\tau_3+\tau_4)\\
    \end{bmatrix}.
\end{equation}
Ideally, the delays $\tau_g$ could be tuned freely, allowing $\xi_{\mu\mu}$ to span the full range of $[-1, 2]$. However, the finite pulse width $\tau_{\rm p}$ imposes practical constraints, as each $\tau_g$ (defined as the time between midpoints of adjacent pulses) must be greater than $\tau_{\rm p}$. This narrows the achievable range of $\xi_{\mu\mu}$. In our experiment, with $\tau_{\rm p}=2\ \mu\text{s}$ and a cycle period of $T\equiv\sum_{g=1}^5\tau_g=30\ \mu\text{s}$, we define a time unit $\tau=T/6=5\ \mu\text{s}$. The resulting constraints on the delays ($0.4\tau \le \tau_g \le 4.4\tau$) leads to ranges of $-0.8 \le \xi_{xx} \le 1.2$ and $-0.6 \le \xi_{yy,zz} \le 1.4$.

To generate off-diagonal terms in the tensor, a non-zero phase $\theta$ is used. This rotates the coordinate system about the $z$-axis, transforming the anisotropy tensor via a similarity transformation, $\bm{\xi}' = R \bm{\xi} R^T$, where $R$ is the standard 2D rotation matrix:
\begin{equation}
    R=
    \begin{bmatrix}
    \cos\theta & -\sin\theta & 0\\
    \sin\theta & \cos\theta & 0\\
    0 & 0 & 1\\
    \end{bmatrix}.
\end{equation}
This transformation populates the off-diagonal $xy$ block:
\begin{align}
    \xi_{xx}'&=\xi_{xx}\cos^2\theta+\xi_{yy}\sin^2\theta,\\
    \xi_{xy}'&=\xi_{yx}'=(\xi_{xx}-\xi_{yy})\sin\theta\cos\theta,\\
    \xi_{yy}'&=\xi_{xx}\sin^2\theta+\xi_{yy}\cos^2\theta.
\end{align}
The $z$-related off-diagonal terms remain zero. To generate those terms (e.g., $\xi_{xz}, \xi_{yz}$), the phase-shifting operators in Eq.~\eqref{eq:seq2} can be replaced with rotations about the $x$- or $y$-axes (e.g., $\mathrm{e}^{\pm\mathrm{i}\hat{O}_x\theta}$).

We engineered three distinct chaotic Hamiltonians, with the parameters specified in \ref{tab:pulse_paras}. The sign of the leading-order Hamiltonian, $\hat{H}_0$, can be effectively reversed by choosing a different set of pulse parameters, allowing for the study of time-reversal dynamics. The parameters used to generate these corresponding time-reversed Hamiltonians ($-\hat{H}_0$) are provided in \ref{tab:pulse_paras_reverse}.
\begin{table}[htb]
    \centering
    \begin{tabular}{c c c c c}
        \toprule
        $\tau_1=\tau_5$ & $\tau_2=\tau_4$ & $\tau_3$ & $\theta$ & Resulting Hamiltonian $\hat{H}_0$\\
        \midrule
        $\tau$ & $0.6\tau$ & $2.8\tau$ & $0$ & $\hat{H}_0^{(1)}\equiv\sum J_{ij}(0.4\hat{S}_{im}^x \hat{S}_{jn}^x-0.4\hat{S}_{im}^y \hat{S}_{jn}^y)$\\
        $\tau$ & $0.6\tau$ & $2.8\tau$ & $\pi/4$ & $\hat{H}_0^{(2)}\equiv\sum J_{ij}(0.4\hat{S}_{im}^x \hat{S}_{jn}^y+0.4\hat{S}_{im}^y \hat{S}_{jn}^x)$\\
        $1.025\tau$ & $0.775\tau$ & $2.4\tau$ & $0$ & $\hat{H}_0^{(3)}\equiv\sum J_{ij}(0.2\hat{S}_{im}^x \hat{S}_{jn}^x-0.225\hat{S}_{im}^y \hat{S}_{jn}^y+0.025\hat{S}_{im}^z \hat{S}_{jn}^z)$\\
        \bottomrule
    \end{tabular}
    \caption{Experimental parameters used to engineer the target Hamiltonians $\hat{H}_0$, with $\tau=5\ \mu\text{s}$.}
    \label{tab:pulse_paras}
\end{table}
\begin{table}[htb]
    \centering
    \begin{tabular}{c c c c c}
        \toprule
        $\tau_1=\tau_5$ & $\tau_2=\tau_4$ & $\tau_3$ & $\theta$ & Resulting Hamiltonian $-\hat{H}_0$ \\
        \midrule
        $\tau$ & $1.4\tau$ & $1.2\tau$ & $0$ & $-\hat{H}_0^{(1)}$\\
        $\tau$ & $1.4\tau$ & $1.2\tau$ & $\pi/4$ & $-\hat{H}_0^{(2)}$\\
        $0.975\tau$ & $1.225\tau$ & $1.6\tau$ & $0$ & $-\hat{H}_0^{(3)}$\\
        \bottomrule
    \end{tabular}
    \caption{Experimental parameters used to engineer the time-reversed Hamiltonians ($-\hat{H}_0$).}
    \label{tab:pulse_paras_reverse}
\end{table}

\subsection{The Error Hamiltonian $\delta\hat{H}$}

The total effective Hamiltonian, $\hat{H}=\hat{H}_0+\hat{H}_1$, includes higher-order correction terms incorporated in $\hat{H}_1$. A perfect reverse of the whole Hamiltonian $\hat{H}$ is physically unrealizable, because on one hand, any real-world control has limitations, such as finite pulse duration, flip-angle error, field inhomogeneity and fluctuation, etc. On the other hand, from a theoretical standpoint, while the sign of $\hat{H}_0$ can be inverted by design, the higher-order Floquet-Magnus terms in $\hat{H}_1$ are generally not reversible. Consequently, the backward evolution is governed by a Hamiltonian $-\hat{H}_0-\hat{H}_1'$, where the deviation from ideal time-reversal is captured by the error Hamiltonian $\delta\hat{H}\equiv\hat{H}_1'-\hat{H}_1$. 

To analyze the components of $\delta\hat{H}$, we performed numerical simulations by calculating the Floquet-Magnus expansion up to the second order (see Eq.~\eqref{eq:FM_expansion}). The simulations were performed on a model of $N=8$ interacting spins with the Hamiltonian:
\begin{equation}
    \hat{H}=\sum_{i<j}J_{ij}(3\hat{S}_i^z\hat{S}_j^z-\hat{\bm{S}}_{i}\cdot\hat{\bm{S}}_{j}).
    \label{eq:nume_model}
\end{equation}
The coupling coefficients $\{J_{ij}\}$ were randomly sampled from a Gaussian distribution $\mathcal{N}(0, \sigma^2)$, with $\sigma = 2J/\sqrt{N}$ and $J=2\pi\times 1460$ Hz. All results were averaged over 100 random realizations of $\{J_{ij}\}$. 

\begin{figure}[htb]
    \centering
    \includegraphics[width=0.7\linewidth]{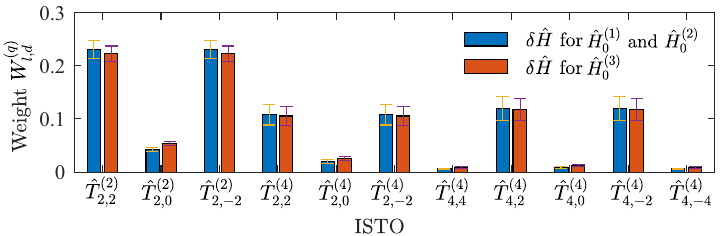}
    \caption{\textbf{Decomposition of the error Hamiltonian $\delta\hat{H}$ into ISTOs.} The weight distribution of $\delta\hat{H}$ in the ISTO basis is shown for the three target Hamiltonians from \ref{tab:pulse_paras}. Simulations were performed on an $N=8$ random spin model (Eq.~\eqref{eq:nume_model}), averaging over 100 realizations. The error term $\delta\hat{H} = \hat{H}_1' - \hat{H}_1$ was calculated using the Floquet-Magnus expansion to second order, assuming ideal pulses and an average pulse interval of $\tau=5\ \mu$s. The weights are normalized such that their sum over all components is unity. The errorbars denote the standard deviations from the samples of 100 realizations. }
    \label{fig:ISTO_weights}
\end{figure}

Firstly, to characterize the structure of the error Hamiltonian, we decomposed $\delta\hat{H}$ into a basis of irreducible spherical tensor operators (ISTOs), $\{\hat{T}_{l,d,s}^{(q)}\}$~\cite{hubbard1969some,suter1988experimental,van2005spherical,man2013cartesian}. Here, $l$ and $d$ are the rank and order of the operator, $q \ge 1$ indicates that the ISTO consists of $q$-body spin operators, and $s$ is an additional index to account for degeneracy. For the symmetric pulse sequences used here, the first-order term $\hat{H}_{\text{1st}}$ vanishes. The dominant contribution to the correction term is therefore the second-order term from Eq.~\eqref{eq:FM_expansion}, i.e., $\hat{H}_1 \approx \hat{H}_{\text{2nd}}$. The Hamiltonians $\hat{H}_\text{2nd}$ and $\hat{H}_\text{2nd}'$ for the forward and backward pulse sequences, respectively, were calculated assuming ideal pulses to isolate errors originating from higher-order Hamiltonian terms. We then projected the error term $\delta\hat{H}\approx\hat{H}_\text{2nd}'-\hat{H}_\text{2nd}$ onto each basis operator,
\begin{equation}
    A_{l,d,s}^{(q)}=\Tr[(\hat{T}_{l,d,s}^{(q)})^\dagger\delta\hat{H}],
    \label{eq:weight}
\end{equation}
and computed the corresponding weights $W_{l,d}^{(q)}=\sum_s|A_{l,d,s}^{(q)}|^2$. The resulting weight distributions for the three typical Hamiltonians are plotted in \ref{fig:ISTO_weights}. The decomposition reveals that $\delta\hat{H}$ is composed predominantly of two-body and four-body operators. The weight distributions for the first two Hamiltonians ($\hat{H}_0^{(1)}$ and $\hat{H}_0^{(2)}$) are identical; this is because their respective pulse sequences differ only by the global rotation $\mathrm{e}^{-\mathrm{i}\hat{O}_z\theta}$, which acts as a unitary transformation on $\delta\hat{H}$ and thus preserves its operator norms.

\begin{figure}[htb]
    \centering
    \includegraphics[width=0.7\linewidth]{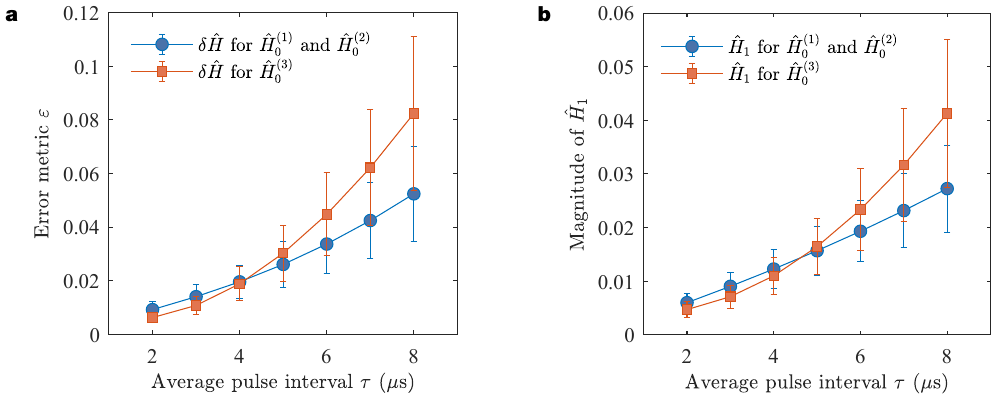}
    \caption{\textbf{Magnitudes of $\delta\hat{H}$ and $\hat{H}_1$ relative to $\hat{H}_0$.} Simulations were performed on an $N=8$ random spin model (Eq.~\eqref{eq:nume_model}), averaging over 100 realizations. The higher-order correction term $\hat{H}_1$ is approximated by the second-order Floquet-Magnus term $\hat{H}_{\rm 2nd}$, and the error Hamiltonian $\delta\hat{H}\approx\hat{H}_\text{2nd}'-\hat{H}_\text{2nd}$. We took a finite pulse width of $\tau_{\rm p}=2\ \mu$s and an average pulse interval $\tau=5\ \mu$s. The errorbars denote the standard deviations from the samples of 100 realizations. }
    \label{fig:magnitude_deltaH}
\end{figure}

To quantify the magnitude of $\delta\hat{H}$, we define a error metric $\varepsilon$:
\begin{equation}
    \varepsilon\equiv\frac{\|\delta\hat{H}\|}{\|\hat{H}_0\|}=\sqrt{\frac{\Tr[(\delta\hat{H})^2]}{\Tr[(\hat{H}_0)^2]}}.
\end{equation}
This calculation incorporates the effect of a finite pulse duration, $\tau_{\rm p}=2\ \mu\text{s}$. As depicted in \ref{fig:magnitude_deltaH}a, $\varepsilon$ increases with the pulse interval $\tau$. For the interval used in our experiment, $\tau=5\ \mu\text{s}$, the calculated errors are $\varepsilon=(2.6\pm 0.9)\%$ for both $\hat{H}_0^{(1)}$ and $\hat{H}_0^{(2)}$, and $\varepsilon=(3.0\pm 1.1)\%$ for $\hat{H}_0^{(3)}$. Furthermore, \ref{fig:magnitude_deltaH}b presents the magnitude of $\hat{H}_1$ relative to $\hat{H}_0$
\begin{equation}
    \frac{\|\hat{H}_1\|}{\|\hat{H}_0\|}\approx\sqrt{\frac{\Tr[(\hat{H}_{\text{2nd}})^2]}{\Tr[(\hat{H}_0)^2]}}.
\end{equation}

\section{Scramblon field theory and its application}

\subsection{Brief introduction to the scramblon field theory}

Scramblon field theory provides an analytical framework to derive various generalizations of out-of-time-order correlators~\cite{KitaevRelationMagnitudeExponent2019,ZhangTwowayApproachOutoftimeorder2022}. It considers the scramblon mode as a novel collective mode, and we ultimately demonstrate that our measured observable can be represented by multiple contributions from scramblon diagrams.

The basic ingredient of scramblon field theory determines the structure of the OTOC. For a general OTOC at infinite temperature, it reads
\begin{equation}
    \label{eq:OTOC1}
    \begin{aligned}
    F_{\hat{W},\hat{V}}
    &=\langle \hat{W}^\dagger(t_1) \hat{V}^\dagger(t_3)\hat{W}(t_2) \hat{V}(t_4) \rangle \\ &=
    \begin{tikzpicture}[scale=0.8]
    \node[ssvertex] (R) at (-5pt,0pt) {};
    \node[ssvertex] (A) at (5pt,0pt) {};
   \draw[] (R)  node[above]{\scriptsize$\hat{W}$};
   \draw[] (A)  node[above]{\scriptsize$\hat{V}$};
    \draw[thick] (R) -- ++(135:10pt) node[left]{\scriptsize$1$};
    \draw[thick] (R) -- ++(-135:10pt) node[left]{\scriptsize$2$};
    \draw[thick] (A) -- ++(45:10pt) node[right]{\scriptsize$3$};
    \draw[thick] (A) -- ++(-45:10pt) node[right]{\scriptsize$4$};
    \end{tikzpicture}
    +\begin{tikzpicture}[scale=0.8]
    \node[ssvertex] (R) at (-13pt,0pt) {};
    \node[ssvertex] (A) at (13pt,0pt) {};
    \draw[] (R)  node[above]{\scriptsize$\hat{W}$};
    \draw[] (A)  node[above]{\scriptsize$\hat{V}$};
    \draw[thick] (R) -- ++(135:10pt) node[left]{\scriptsize$1$};
    \draw[thick] (R) -- ++(-135:10pt) node[left]{\scriptsize$2$};
    \draw[thick] (A) -- ++(45:10pt) node[right]{\scriptsize$3$};
    \draw[thick] (A) -- ++(-45:10pt) node[right]{\scriptsize$4$};
    \draw[wavy] (A) to (R);
    \end{tikzpicture}
    +
    \begin{tikzpicture}[scale=0.8]
    \node[ssvertex] (R) at (-13pt,0pt) {};
    \node[ssvertex] (A) at (13pt,0pt) {};
    \draw[] (R)  node[above]{\scriptsize$\hat{W}$};
    \draw[] (A)  node[above]{\scriptsize$\hat{V}$};
    \draw[thick] (R) -- ++(135:10pt) node[left]{\scriptsize$1$};
    \draw[thick] (R) -- ++(-135:10pt) node[left]{\scriptsize$2$};
    \draw[thick] (A) -- ++(45:10pt) node[right]{\scriptsize$3$};
    \draw[thick] (A) -- ++(-45:10pt) node[right]{\scriptsize$4$};
    \draw[wavy] (A) to[out=140,in=40] (R);
    \draw[wavy] (A) to[out=-140,in=-40] (R);
    \end{tikzpicture}+\cdots
    \\
&=\sum_{k=0}^{\infty}\frac{(-\lambda)^{k}}{k!}\VF^{\R,k}_{\hat{W}}(t_1-t_2)\VF^{\A,k}_{\hat{V}}(t_3-t_4).
\end{aligned}
\end{equation}
To ensure the scramblon enters the free collective modes region, we ensure $t_1 \approx t_2 \gg t_3 \approx t_4$. Then $\lambda = C^{-1} e^{\varkappa \frac{t_1+t_2-t_3-t_4}{2}}$ is the propagator for the scrambling modes, which are depicted by the wavy lines. The prefactor $C$ is proportional to the number of degrees of freedom $N$, with a coefficient related to the Lyapunov exponent and the branching time \cite{KitaevRelationMagnitudeExponent2019,KitaevObstacleSubAdSHolography2021}. Since we consider infinite temperature, all phase factors related to the chaos bound are suppressed \cite{StanfordBoundChaos2016}. The vertices $\VF^{\R,k}$ and $\VF^{\A,k}$, represented by the gray dots, are the scattering vertices between operators $\hat{W}, \hat{V}$ and $k$ scramblons in the future and the past, respectively. For a unified description, we need to sum over all $k$ to describe both the early-time and near-scrambling-times regions~\cite{KitaevRelationMagnitudeExponent2019,ZhangTwowayApproachOutoftimeorder2022}. However, we can revert to the early-time limit by only considering $k=0,1$, as $\lambda \approx e^{\varkappa t}/N \ll 1$ at early times.

\subsection{Derivation of the Correlators}
As discussed in the main text, there are two types of OTOCs involved in the calculation of the correlator $F_{\alpha\gamma}(\phi,t)$:
\begin{equation}\label{eqn:reduction1}
F_{\alpha \gamma}^{(0)}(\phi,t)=\text{Tr}[ e^{-i\phi\hat{O}_\alpha (t)}\hat{O}_\gamma(0)e^{i\phi\hat{O}_\alpha (t)} \hat{O}_\gamma(0)]/\mathcal{C}\ \ \ \ \text{Type 1},
\end{equation}
\begin{equation}\label{eqn:reduction2}
F_{\alpha \gamma}(0,t)=\text{Tr}[ \hat{O}_\gamma(0)\hat{\mathcal{V}}^\dagger(t) \hat{O}_\gamma(0)\hat{\mathcal{V}}(t)]/\mathcal{C}\ \ \ \ \text{Type 2},
\end{equation}
By definition, the most elementary building blocks are the OTOCs for Type 1 ($\hat{W}=\hat{O}_{\alpha}, \hat{V}=\hat{O}_{\gamma}$) and Type 2 ($\hat{W}=\delta\hat{H}, \hat{V}=\hat{O}_{\gamma}$), for which well-defined ans\"atze for the vertex functions $\VF^{\R/\A}_{\hat{O}_{\alpha}}$ and $\VF^{\R/\A}_{\delta \hat{H}}$ exist. After handling the exponential in operator $\hat{W}$ by expansion, we expect to obtain the diagrammatic contributions to the correlator $F_{\alpha\gamma}(\phi,t)$ using only the ansatz of vertex functions and scramblon modes.
Within the early-time approximation, we consider the diagrams as follows:
\begin{equation}
   \includegraphics[width=0.7\linewidth]{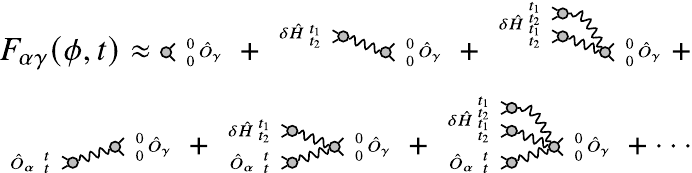}
\end{equation}
These diagrams can be summarized by the equation:
\begin{equation}
    \begin{aligned}\label{eq:Fphi_diagrams}
        F_{\alpha\gamma}(\phi,t)&= \sum_{l_1,l_2=0}^{\infty}  \frac{\left(-\int_{0}^{t}\int_{0}^{t}\diff{t_1}\diff{t_2} \lambda_{1} \VF^{\R,1}_{\delta\hat{H}}(t_1-t_2)\right)^{l_1}}{l_1!} \frac{\left(-\lambda_2\phi^2 \VF^{\R,1}_{\hat{O}_{\alpha}}(0)\right)^{l_2}}{l_2!} \VF^{\A,l_1+l_2}_{\hat{O}_{\gamma}}(0) / \mathcal{C}.
    \end{aligned}
\end{equation}
The scramblon propagators depend on the time variables of the operators, specifically $\lambda_{1}=C^{-1}e^{\varkappa(t_1+t_2)/2}$ and $\lambda_{2}=C^{-1}e^{\varkappa t}$, respectively. In this equation, we only consider the early-time contribution to the scramblon; practically, this leads to a simpler ansatz in this region, and physically, time-reversal imperfection naturally suppresses late-time dynamics. Therefore, only up to one scramblon can be attached to the vertex on the left-hand side. Additionally, due to the exponential structure in operators, the zeroth-order vertex functions for $\delta\hat{H}, \hat{S}_{\alpha}$, defined as the two-point correlation function $\VF^{\R,0}_{\hat{O}}(t_1,t_2) \equiv \langle \hat{O}(t_1) \hat{O}(t_2) \rangle$, ultimately cancel out. Hence, only the vertex function $\VF^{\R,1}_{\hat{O}}$ appears in Eq.~\eqref{eq:Fphi_diagrams}.

Using the techniques developed in scramblon theory \cite{KitaevRelationMagnitudeExponent2019,ZhangTwowayApproachOutoftimeorder2022}, we represent the vertex functions in terms of two new functions, for which analytical expressions are known for some solvable systems. The vertex functions are the moments of the distributions $h^{\R/\A}$:
      \begin{equation}
          \VF^{\R/\A,k}(t) = \int_0^{+\infty} y^k h^{\R/\A} (y,t) dy.
      \end{equation}
Then, we introduce:
\begin{equation}
\label{eqn: fh}
f^{\R/\A}(x,t) = \sum_{k=0}^\infty \frac{(-x)^k}{k!} \VF^{\R/\A,k}(t)=\int_0^{+\infty} e^{-xy} h^{\R/\A} (y,t) dy .
\end{equation}
The functions $f^{\R/\A}(x,t)$ can be interpreted as correlation functions on the perturbed background of the thermofield double. The parameter $x$ characterizes the strength of the mean-field perturbation. Through this parametrization, we find that a general OTOC like Eq.~\eqref{eq:OTOC1} can have a similar form to that in the black hole problem \cite{shenkerStringyEffectsScrambling2015}, where variables $y$ are analogous to the null momenta on the past/future horizon.

By expanding the vertex function and rearranging terms, we can obtain the final result for the correlator:
\begin{equation}\label{eq:F_simplify_result}
    \begin{split}
   F_{\alpha \gamma}(\phi, t)
    &= \int_{0}^{+\infty}\diff{y} \sum_{l_1,l_2=0}^{\infty}  \frac{\left(-y\int_{0}^{t}\int_{0}^{t}\diff{t_1}\diff{t_2} \lambda_{1} \VF^{\R,1}_{\delta\hat{H}}(t_1-t_2)\right)^{l_1}}{l_1!} \frac{\left(-y\lambda_2\phi^2 \VF^{\R,1}_{\hat{O}_{\alpha}}(0)\right)^{l_2}}{l_2!} h^{\A}_{\hat{O}_{\gamma}}(0) / \mathcal{C} \\
    &= \int_{0}^{+\infty}\diff{y}    \exp\left(-y \left(\int_{0}^{t}\int_{0}^{t}\diff{t_1}\diff{t_2} \lambda_{1} \VF^{\R,1}_{\delta\hat{H}}(t_1-t_2)+\lambda_2\phi^2 \VF^{\R,1}_{\hat{O}_{\alpha}}(0)\right) \right) h^{\A}_{\hat{O}_{\gamma}}(0) / \mathcal{C} \\
    &= f^{\A}_{\hat{O}_{\gamma}}\left(    \int_{0}^{t}\int_{0}^{t}\diff{t_1}\diff{t_2} \lambda_{1} \VF^{\R,1}_{\delta\hat{H}}(t_1-t_2)+\lambda_2\phi^2 \VF^{\R,1}_{\hat{O}_{\alpha}}(0), 0\right) / \mathcal{C}. \\
    \end{split}
\end{equation}
This is the central result of the scramblon theory. The vertex function of global operators is $\mathcal{O}(N)$, which, when combined with $C^{-1}\propto N^{-1}$, leads to an $\mathcal{O}(1)$ value. The $N$ factor in the function $f^{\A}_{\hat{O}_{\gamma}}$ of the global spin operator is also cancelled by the normalization constant $\mathcal{C}$. Therefore, the scramblon theory leads to a reasonable prediction in the thermodynamical limit $N\rightarrow \infty$.

\subsubsection{General Proof of Eq.~\eqref{eq:F_simplify_result}}

First, we calculate an intermediate result. We consider the OTOC between $e^{\iu \hat{W}}$ and $\hat{V}$, which can be expanded as:
\begin{equation}\label{eq:F_earlytime}
    \begin{split}
        F_1&=\langle e^{\iu \hat{W}(T_1)}\hat{V}e^{-\iu \hat{W}(T_2)}\hat{V} \rangle/\mathcal{C} \\
        &= \sum_{l=0}^{\infty}\sum_{s=0}^{l} \frac{(-1)^s f(s)}{(l-s)!(l+s)!} \langle [\hat{W}(t_1)\cdots \hat{W}(t_{l-s})] \hat{V} [\hat{W}(t_{l-s+1})\cdots \hat{W}(t_{2l})] \hat{V}  \rangle /\mathcal{C} \\
    \end{split}
\end{equation}
Here, we introduce $2l$ time variables after expansion to label the index of the operator. For the Type 2 operator, we choose the convention $t_1, \dots, t_{l-s} = T_1$ and $t_{l-s+1}, \dots, t_{2l} = T_2$.

Using scramblon theory, we pair the operators to form the OTOCs. As a simplest example, we choose all $(l-s)$ operators in the first interaction vertex and pair them with $(l-s)$ operators in the second interaction vertex. This will lead to $C_{l+s}^{l-s}$ choices. As a result, two sets of $(l-s)$ $\hat{W}(t_w)$ effectively pair to form the function $\prod_{w=1}^{l-s} \VF_{\hat{W}}^{\R,k}(\lambda y, (t_{1w}-t_{2w}))$, with $t_{1w} \in \{t_1, \dots, t_{l-s}\}$ and $t_{2w} \in \{t_{l-s+1}, \dots, t_{2l}\}$, along with an extra combination number $(l-s)!$ by fixing $t_{1w}$ to be close to $t_{2w}$. Apart from this, the $2s$ $\hat{W}(t_w)$ operators in the right interaction vertex will be factored out of the OTOCs and contribute to $\prod_{v=0}^{s} G_{\hat{W}}(t_{1v} - t_{2v})$, where $t_{1v}, t_{2v} \in \{t_{l-s+1}, \dots, t_{2l}\} \setminus \{t_{2w}\}$. By applying Wick's theorem, this also contributes a factor $(2s-1)!!$.
Finally, it leads to one specific contribution:
\begin{equation}
    \begin{split}
        &\sum_{l=0}^{\infty} \sum_{s=0}^{l} \frac{\eta^{2l} (-1)^s f(s)}{(l-s)!(l+s)!} C_{l+s}^{l-s} (l-s)! (2s-1)!!    \sum_{\{k_w\}} \prod_{w=1}^{l-s} \left( \frac{(-\lambda_w)^{k_w}}{k_w}    \VF^{R,k_w}_{\hat{W}}(t_{1w}-t_{2w})\right) \prod_{v=0}^{s} \left( G_{\hat{W}}(t_{1v} - t_{2v}) \right)\VF_{\hat{V}}^{\A,\sum_w k_w}(0) \\
    =&  \int \diff{y}\, h^R_{\hat{V}}(y,0) \sum_{l=0}^{\infty} \sum_{s=0}^{l} \frac{\eta^{2l} (-1)^s f(s)}{(l-s)!(l+s)!} C_{l+s}^{l-s} (l-s)! (2s-1)!!    \prod_{w=1}^{l-s} \left(  f^{\R}_{\hat{W}}(\lambda_w y, (t_{1w}-t_{2w}))\right) \prod_{v=0}^{s} \left( G_{\hat{W}}(t_{1v} - t_{2v}) \right), \\
    \end{split}
\end{equation}
where the scramblon propagator is the same for these specific time variables: $\lambda_w = \lambda = C^{-1}e^{\varkappa (T_1+T_2)/2}$.

\begin{figure}[htb]
    \centering
    \includegraphics[width=0.6\linewidth]{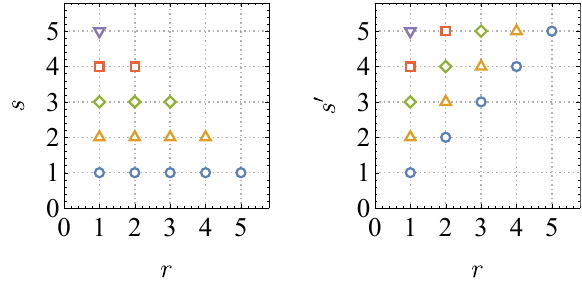}
    \caption{\textbf{The mapping of the index $(s,r)$ to $(s',r)$}. The mapping of the same $s$ is plotted with the same markers.}
    \label{fig:supp_mapping}
\end{figure}

However, starting from the previous case, we notice that $(l-s)$ operators of $\hat{W}$ on either the left or right side can be self-contracted using Wick's theorem. For example, the left interaction vertex has contracted $2r$ operators with $r=1,2,\dots,\floor*{(l-s)/2}$, and therefore we have $l-s'=l-(s+2r)$ OTOCs to pair. Similarly, we have the combination numbers $C_{l-s}^{l-(s+2r)}$ and $C_{l+s}^{l-(s+2r)}$, which arise from the choice of $l-(s+2r)$ scramblons from the $(l-s)$ operators or $(l+s)$ operators. We further have the $(2r-1)!!$ and $(2(r+s)-1)!!$ factors when applying Wick's theorem at the left- and right-hand sides to obtain the Green's function $G$. The full expression reads:
    \begin{equation}
        \begin{split}
            F_1
            &= \sum_{l=0}^{\infty}\sum_{s=0}^{l}\sum_{r=0}^{\floor*{(l-s)/2}} \frac{(-1)^s f(s)}{(l-s)!(l+s)!} C_{l-s}^{l-(s+2r)} C_{l+s}^{l-(s+2r)} (2r-1)!! (2(r+s)-1)!! \\
            &\qquad \sum_{\{k_w\}} \prod_{w=1}^{l-(s+2r)} \left( \frac{(-\lambda_w)^{k_w}}{k_w}    \VF^{R,k_w}_{\hat{W}}(t_{1w}-t_{2w})\right) \prod_{v=0}^{s+2r} \left( G_{\hat{W}}(t_{1v} - t_{2v}) \right)\VF_{\hat{V}}^{\A,\sum_w k_w}(0) /\mathcal{C} \\
            &= \int \diff{y}\, h^{\A}_{\hat{V}}(y,0)\sum_{l=0}^{\infty}\sum_{s=0}^{l}\sum_{r=0}^{\floor*{(l-s)/2}} \frac{(-1)^s f(s)}{(l-s)!(l+s)!} C_{l-s}^{l-(s+2r)} C_{l+s}^{l-(s+2r)} (2r-1)!! (2(r+s)-1)!! \\
            &\qquad \left(  f^{\R}_{\hat{W}}(\lambda y, (t_{3}-t_{4}))\right)^{l-(s+2r)}  \left(G_{\hat{W}}(0) \right)^{s+2r} /\mathcal{C} \\
        \end{split}
    \end{equation}
First of all, we can replace $s'=s+2r$, and then the summation $\sum_{s=0}^{l}\sum_{r=0}^{\floor*{(l-s)/2}} \to \sum_{s'=0}^{l}\sum_{r=0}^{\floor*{s'/2}}$, which is because $s=s'-2r \geq 0$. The mapping is illustrated in \ref{fig:supp_mapping}.
Consequently, we find the coefficient can be simplified as:
    \begin{equation*}
        \begin{split}
            &\sum_{s=0}^{l}\sum_{r=0}^{\floor*{(l-s)/2}} \frac{\eta^{2l} (-1)^s f(s)}{(l-s)!(l+s)!} C_{l-s}^{l-(s+2r)} C_{l+s}^{l-(s+2r)} (2r-1)!! (2(r+s)-1)!! \\
        = & \sum_{s'=0}^{l}\sum_{r=0}^{\floor*{s'/2}} (-1)^{s'}\frac{1}{(l-s')!}\frac{f(s'-2r)}{r!(s'-r)! 2^{s'}}
        \end{split}
    \end{equation*}
We can perform the summation over $r$ first, because the exponent of $f^A_{\hat{W}}$ and $G_{\hat{W}}$ is independent of $r$. When $s'$ is even, we have $\sum_{r=0}^{\floor*{s'/2}} \frac{f(s'-2r)}{r!(s'-r)!} s'! = (2C^0_{s'}+2C^1_{s'}+\cdots+2C^{s'/2-1}_{s'}+C^{s'/2}_{s'}) = 2^{s'}$. Similarly, when $s'$ is odd, we have $\sum_{r=0}^{\floor*{s'/2}} \frac{f(s'-2r)}{r!(s'-r)!} s'!= (2C^0_{s'}+2C^1_{s'}+\cdots+2C^{(s'-1)/2}_{s'}) = 2^{s'}$. Therefore, after summation, $\sum_{s'=0}^{l}\sum_{r=0}^{\floor*{s'/2}} (-1)^{s'}\frac{1}{(l-s')!}\frac{f(s'-2r)}{r!(s'-r)! 2^{s'}} = \sum_{s'=0}^{l} \frac{(-1)^{s'}}{l!} C_{l}^{s'}$.
    \begin{equation}\label{eq:OTOC_expOp}
        \begin{split}
            F_1(t) &= \int \diff{y}\, h^{\A}_{\hat{V}}(y,0) \sum_{l=0}^{\infty}\sum_{s'=0}^{l}\frac{1}{l!}C_{l}^{s'}\left(f^{\R}_{\hat{W}}(\lambda y, (T_1-T_2))\right)^{l-s'}  \left(- G_{\hat{W}}(T_1-T_2) \right)^{s'} /\mathcal{C} \\
            &= \int \diff{y}\, h^{\A}_{\hat{V}}(y,0) \exp\left(f^{\R}_{\hat{W}}(\lambda y, (T_1-T_2)) - G_{\hat{W}}(T_1-T_2) \right) /\mathcal{C},\\
        \end{split}
    \end{equation}
where $\lambda = C^{-1}e^{\varkappa (T_1+T_2)/2}$ is the scramblon propagator.

We apply the result of Eq.~\eqref{eq:OTOC_expOp} to the case of two operators, with Type 1 being $\hat{W}=\hat{O}_{\alpha}, \hat{V}=\hat{O}_{\gamma}$, and Type 2 being $\hat{W}=\int \diff t'\delta\hat{H}(t'), \hat{V}=\hat{O}_{\gamma}$. Using the same procedure, the final result reads:
\boxalign{\begin{equation}\label{eq:F_scramblon_alltime}
    F_{\alpha\gamma}(\phi,t) = \int \diff{y}\, h^{\A}_{\hat{O}_{\gamma}}(y,0) \exp\left(    \int_{0}^{t}\int_{0}^{t}\diff{t_1}\diff{t_2} \left(f^{\R}_{\delta\hat{H}}(\lambda_1 y, (t_1-t_2)) - G_{\delta\hat{H}}(t_1-t_2)\right) + \phi^2\left(f^{\R}_{\hat{O}_{\alpha}}(\lambda_2\phi^2 y, 0) - G_{\hat{O}_{\alpha}}(0)\right) \right) / \mathcal{C}.
\end{equation}
}
Equation~\eqref{eq:F_scramblon_alltime} is the central result derived from the scramblon theory. Here, we do not assume the early-time approximation, but only the conditions $\phi\ll 1$ and $\delta H/H \ll 1$. This is because, with a fixed coupling constant $\phi$ in the large $N$ limit, each term $\hat{O}^k$ from expanding the exponential $e^{\pm i \phi \hat{O}}$ will carry an $N^k$ factor due to the summation of the local operators. This factor cancels the $N^{-k}$ factor originating from the leading ladder diagrams. Consequently, the leading diagrams in the large $N$ limit are already of order $1$ without amplification from chaotic time evolutions. This implies that the correlator $F_{\alpha\gamma}(\phi,t)$ is already zero initially. Therefore, we require $\phi \ll 1$, which allows us to safely replace each ladder with a propagator of the scrambling mode.

With the early-time approximation, we can use $f^{\R}_{\hat{O}_{\alpha}}(\lambda y, t) - G_{\hat{O}_{\alpha}}(t) \approx -\lambda y \VF_{\hat{O}_{\alpha}}^{\R,1}(t)$, which immediately leads to the result of Eq.~\eqref{eq:F_simplify_result}.

\subsection{Simplification of the Correlators}

Based on the result of Eq.~\eqref{eq:F_simplify_result}, we simplify the result with some approximations to obtain a simpler ansatz. The scramblon propagators depend on the time variables of the operators; specifically, we have $\lambda_{1}=C^{-1}e^{\varkappa(t_1+t_2)/2}$ and $\lambda_{2}=C^{-1}e^{\varkappa t}$. For the first term, we can further simplify the result by separating the time variables into the center of mass $t_c\equiv(t_1+t_2)/2$ and the relative time difference $t'=t_1-t_2$. We use the assumption here to modify $\int_{0}^{t}\int_{0}^{t}\diff{t_1}\diff{t_2} \approx \int_{0}^{t}\diff{t_c} \int_{-\infty}^{\infty}\diff{t'}$, which then reduces the first term in Eq.~\eqref{eq:F_earlytime} to $-C^{-1}\varkappa^{-1}(e^{\varkappa t}-1) \tVF^{\R,1}_{\delta\hat{H}}$, where the constant is $\tVF^{\R,1}_{\delta\hat{H}} = \int_{-\infty}^{\infty} \diff{t'} \VF^{\R,1}_{\delta\hat{H}}(t')$. We find that only the $t$-dependence enters through $(e^{\varkappa t}-1)$, while other terms are model-dependent constants.

In general, it is challenging to obtain an analytical formula for $f^{\R}_{\hat{O}_{\gamma}}(x,t)$ in randomly interacting spin systems. However, all known solvable models, such as large-$q$ SYK, Brownian SYK, or Brownian circuits, exhibit a similar ansatz:
\begin{equation}\label{eq:f_ansatz}
    f^R_{\hat{O}_{\gamma}}(x,0) = f^A_{\hat{O}_{\gamma}}(x,0) = \mathcal{C} \left(1 + c_{\gamma} x\right)^{-2\Delta_{\gamma}},
\end{equation}
where the parameters $c_{\gamma}$ and the scaling dimension $\Delta_{\gamma}$ depend on both the choice of model and the detection direction $\gamma$. For each $\alpha, \gamma = x,y,z$, the data are fitted by the equation
\begin{equation}\label{eqn:main_scramblon}
F_{\alpha \gamma }(\phi,t)= \frac{1}{\left(1+a_\gamma e^{\varkappa t}+b_{\alpha \gamma} \phi^2 e^{\varkappa t}\right)^{2\Delta_\gamma}}.
\end{equation}
in the main text. Here, we explicitly illustrate the direction dependence of $a_{\gamma}$, $b_{\alpha \gamma}$, and $\Delta_{\gamma}$. According to both Eq.~\eqref{eq:F_simplify_result} and Eq.~\eqref{eqn:main_scramblon}, two of the three parameters are identified as:
\begin{equation}\label{eq:parameters}
    a_{\gamma} =  c_{\gamma} C^{-1}\varkappa^{-1}\tilde{\VF}^{\R,1}_{\delta \hat{H}}, \ \ \  b_{\alpha \gamma} = \VF^{\R,1}_{\hat{O}_{\alpha}}(0) c_{\gamma}  C^{-1}.
\end{equation}
Besides, the vertex function of the spin operators is related to the ansatz in Eq.~\eqref{eq:f_ansatz} by the relation $\VF^{\R/\A,1}_{\hat{O}_{\gamma}}(t) = -\partial_x f^{\R/\A}_{\hat{O}_{\gamma}}(x,t)|_{x=0}$, which leads to $\VF^{\R/\A,1}_{\hat{O}_{\gamma}} = 2\Delta_{\gamma}c_{\gamma}$.

Then we can verify the three relations satisfied by scramblon theory:
\begin{enumerate}[label=R\arabic*.]
\item Firstly, after simplification, Type 1 and Type 2 scramblon propagators, $\lambda_1$ and $\lambda_2$, still contribute the same scramblon mode, proportional to $e^{\varkappa t}$, which is independent of the vertex functions and, therefore, the choice of operators $\alpha$ and $\gamma$.

\item Secondly, scramblon theory predicts that $M_{\alpha\gamma}=b_{\alpha \gamma} \Delta_\gamma = 2 \Delta_{\alpha}c_{\alpha}\Delta_{\gamma}c_{\gamma}\mathcal{C}C^{-1}$ is a symmetric matrix in $\alpha$ and $\gamma$, which should satisfy $\text{rank}(M_{\alpha\gamma})=1$.

\item Finally, according to the form of ansatz Eq.~\eqref{eq:f_ansatz} and the identification of parameters in Eq.~\eqref{eq:parameters}, we conclude that both $a_\gamma$ and $\Delta_\gamma$ should be independent of $\alpha$. Note that the vertex function $\tilde{\VF}^{\R,1}_{\delta \hat{H}}$ is independent of how we measure the system.
\end{enumerate}

\section{Fitting of the experimental data}

In this section, we summarize the optimal fitting curves in~\ref{fig:optimal_fit} and the fitting parameters in~\ref{tab:data_fitting}. Then we quantify the goodness of the fitting by checking the adjusted-$R^2$ value in \ref{fig:supp_R2adjust}.

\subsection{Fitting parameter error analysis}
When estimating the error in the fitting parameters presented in \ref{tab:data_fitting}, we considered three primary sources.
First, for a \textbf{fixed fitting range}, the best-fit parameters $\vartheta$ are determined with a residual standard deviation, $\sigma_{\vartheta,\text{res}}$. For our data, we found that $\sigma_{\text{res}} / \vartheta \gtrsim 10^{-2}$.
Second, the \textbf{choice of fitting range} significantly influences the best-fit parameters. This is theoretically sound, as the scramblon is expected to become non-interacting only after a certain timescale \cite{ZhangTwowayApproachOutoftimeorder2022}. Experimentally, we varied the initial fitting time step from $2\Delta\tau$ to $10\Delta\tau$, where $\Delta\tau$ is the experimental sampling time step, while keeping the final fitting time step constant at the last sampled time. From these nine sets of best-fit parameters, we calculated their mean $\bar{\vartheta}$ and standard deviation $\sigma_{\vartheta,\text{range}}$. For our data, we observed $\sigma_{\vartheta,\text{range}} / \bar{\vartheta} \gtrsim 10^{-2}$.
Third, the \textbf{readout error} is $10^{-4}$ for an $O(1)$ signal. This indicates that the intrinsic error in the raw data is considerably smaller than the aforementioned two error sources, and thus, we neglect it in our analysis.

Detailed information regarding our error processing methodology can be found in the accompanying Mathematica script, \texttt{draft\_fig3.nb} \cite{code_zenodo}.

\begin{table*}[htb]
    \centering
    \small
    \begin{tabular}{cccccc}
        \toprule
      & $\varkappa$
      & Eig $M^s$ &  Eig $M^a$ & $\Delta$ & $a \times 50$ \\
      \midrule
	\text{Ham 1} & $ \left(
	\begin{array}{ccc}
		\text{0.45(2)} & \text{0.46(3)} & \text{0.42(4)} \\
		\text{0.44(2)} & \text{0.43(2)} & \text{0.42(4)} \\
		\text{0.46(2)} & \text{0.46(3)} & \text{0.44(4)} \\
	\end{array}
	\right) $ & $ \left(
	\begin{array}{c}
		\text{1.5(4)} \\
		\text{0.13(8)} \\
		\text{0.04(3)} \\
	\end{array}
	\right) $ & $ \left(
	\begin{array}{c}
		\text{0.13(8)} \\
		\text{0.041(4)} \\
		\text{0} \\
	\end{array}
	\right) $ & $ \left(
	\begin{array}{ccc}
		\text{0.34(3)} & \text{0.33(4)} & \text{0.6(2)} \\
		\text{0.35(3)} & \text{0.36(3)} & \text{0.6(1)} \\
		\text{0.33(3)} & \text{0.33(4)} & \text{0.6(1)} \\
	\end{array}
	\right) $ & $ \left(
	\begin{array}{ccc}
		\text{0.20(3)} & \text{0.17(3)} & \text{0.26(5)} \\
		\text{0.21(3)} & \text{0.20(3)} & \text{0.25(4)} \\
		\text{0.18(3)} & \text{0.17(4)} & \text{0.23(5)} \\
	\end{array}
	\right)$ \\
    \midrule
	\text{Ham 2} & $ \left(
	\begin{array}{ccc}
		\text{0.47(3)} & \text{0.45(2)} & \text{0.41(4)} \\
		\text{0.44(1)} & \text{0.45(2)} & \text{0.42(4)} \\
		\text{0.47(3)} & \text{0.46(2)} & \text{0.45(4)} \\
	\end{array}
	\right) $ & $ \left(
	\begin{array}{c}
		\text{1.5(4)} \\
		\text{0.17(8)} \\
		\text{0.09(4)} \\
	\end{array}
	\right) $ & $ \left(
	\begin{array}{c}
		\text{0.17(8)} \\
		\text{0.038(7)} \\
		\text{0} \\
	\end{array}
	\right) $ & $ \left(
	\begin{array}{ccc}
		\text{0.33(4)} & \text{0.33(3)} & \text{0.7(2)} \\
		\text{0.36(2)} & \text{0.35(3)} & \text{0.6(2)} \\
		\text{0.32(3)} & \text{0.33(3)} & \text{0.6(1)} \\
	\end{array}
	\right) $ & $ \left(
	\begin{array}{ccc}
		\text{0.18(4)} & \text{0.19(3)} & \text{0.28(5)} \\
		\text{0.24(2)} & \text{0.21(3)} & \text{0.27(5)} \\
		\text{0.19(4)} & \text{0.21(4)} & \text{0.24(5)} \\
	\end{array}
	\right)$ \\
    \midrule
	\text{Ham 3} & $ \left(
	\begin{array}{ccc}
		\text{0.228(7)} & \text{0.247(7)} & \text{0.22(2)} \\
		\text{0.224(6)} & \text{0.228(5)} & \text{0.22(2)} \\
		\text{0.230(8)} & \text{0.236(7)} & \text{0.23(2)} \\
	\end{array}
	\right) $ & $ \left(
	\begin{array}{c}
		\text{1.8(4)} \\
		\text{0.09(8)} \\
		\text{0.03(2)} \\
	\end{array}
	\right) $ & $ \left(
	\begin{array}{c}
		\text{ 0.09(8)} \\
		\text{ 0.016(3)} \\
		\text{ 0} \\
	\end{array}
	\right) $ & $ \left(
	\begin{array}{ccc}
		\text{0.49(4)} & \text{0.24(1)} & \text{0.7(2)} \\
		\text{0.49(3)} & \text{0.26(1)} & \text{0.7(1)} \\
		\text{0.48(4)} & \text{0.25(1)} & \text{0.6(2)} \\
	\end{array}
	\right) $ & $ \left(
	\begin{array}{ccc}
		\text{0.45(2)} & \text{0.45(4)} & \text{0.61(4)} \\
		\text{0.45(2)} & \text{0.56(3)} & \text{0.60(3)} \\
		\text{0.43(3)} & \text{0.51(4)} & \text{0.53(6)} \\
	\end{array}
	\right)$ \\
\bottomrule
    \end{tabular}
    \caption{\textbf{Fitting parameters from the experimental data.} For the parameters $\varkappa$, $\Delta$, and $a$, the rows of the respective matrices correspond to the $\alpha$ direction, and the columns correspond to the $\gamma$ direction. For $\text{Eig } M^s$ or $\text{Eig } M^a$, the three rows represent the absolute value of three eigenvalues. Data for all three Hamiltonians are presented. The numerical value 0 without a standard error indicates machine precision.}
    \label{tab:data_fitting}
\end{table*}

\subsection{Goodness of the fitting}

To further verify our theory, we quantify goodness of fit using the adjusted $R$-squared, denoted $\bar R^{2}$. Specifically, we evaluate $\bar R^{2}$ with Mathematica’s \texttt{NonlinearModelFit[]} and the \texttt{"AdjustedRSquared"} property. Smaller values of
$
1-\bar R^{2}
$
indicate better fits.

In ~\ref{fig:supp_R2adjust} we report $\bar R^{2}$ for all experimental datasets summarized in the Supplemental Material. For the datasets shown in Fig.~2, we obtain
$
1-\bar R^{2}={3\times 10^{-4},\ 1\times 10^{-4},\ 3\times 10^{-4}}.
$
The worst case across all datasets is
$
1-\bar R^{2}=7\times 10^{-4},
$
which is only about twice the values in Fig.~2 in the main text.

The trends in $\bar R^{2}$ are consistent with Fig.~3 in the main text: configurations with $\gamma=z$ exhibit reduced goodness of fit (larger error bars in Fig.~3). This behavior may originate from systematic errors in the experimental setup and warrants further investigation.

\begin{figure}[htb]
\centering
\includegraphics[width=0.4\linewidth]{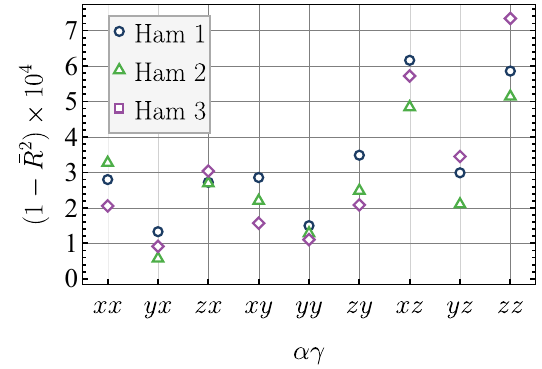}
\caption{Adjusted $R$-squared, $\bar R^{2}$, for all experimental datasets presented in the \ref{fig:optimal_fit}. The horizontal axis labels the spin indices $\alpha\gamma$ of the OTOC $F_{\alpha\gamma}(\phi,t)$. The vertical axis shows $(1-\bar R^{2}) \times 10^4$. Different markers denote the three Hamiltonians implemented in the experiment.}
\label{fig:supp_R2adjust}
\end{figure}

\begin{figure}
    \centering
    \includegraphics[width=0.7\linewidth]{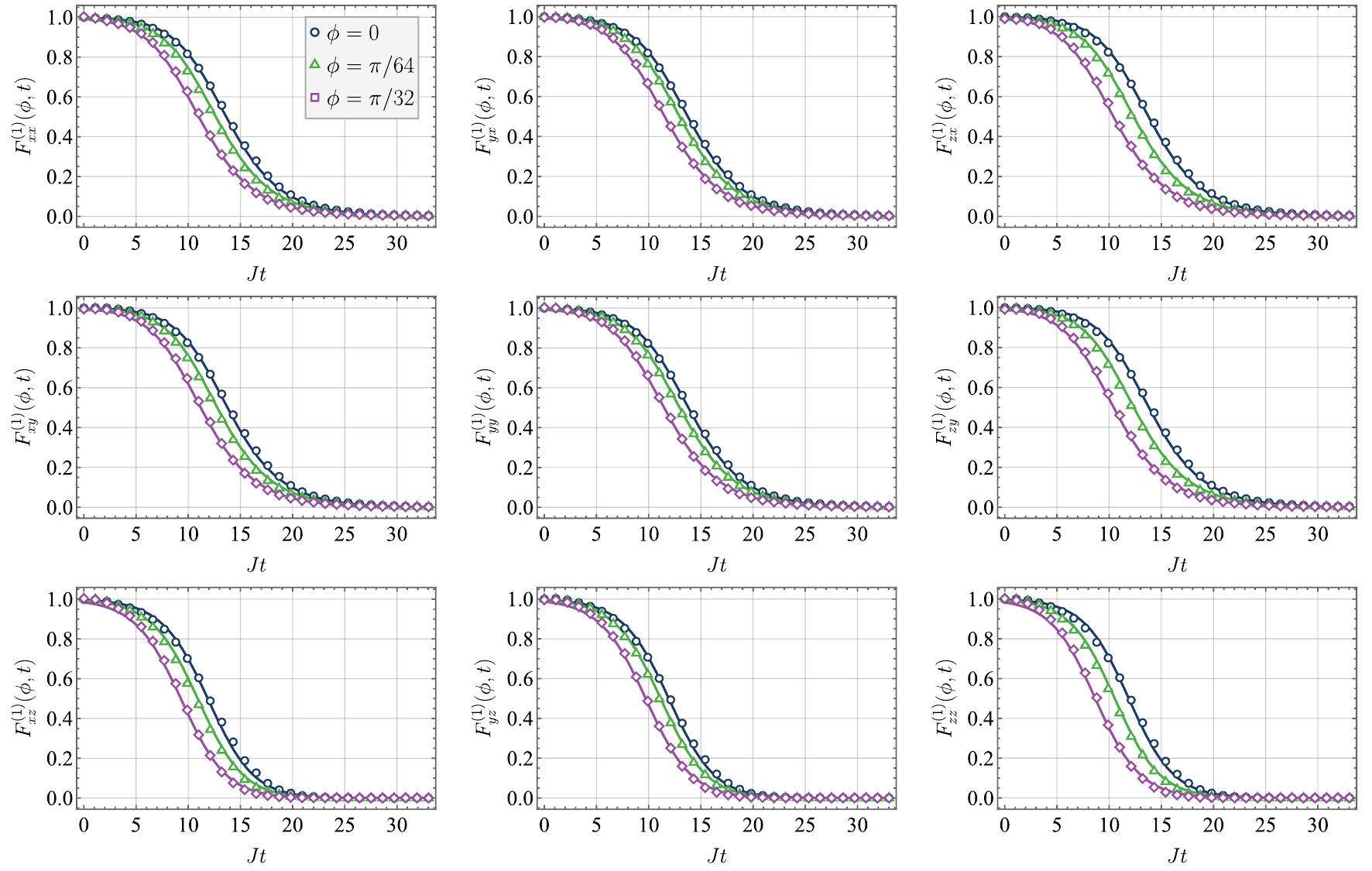}
    \includegraphics[width=0.7\linewidth]{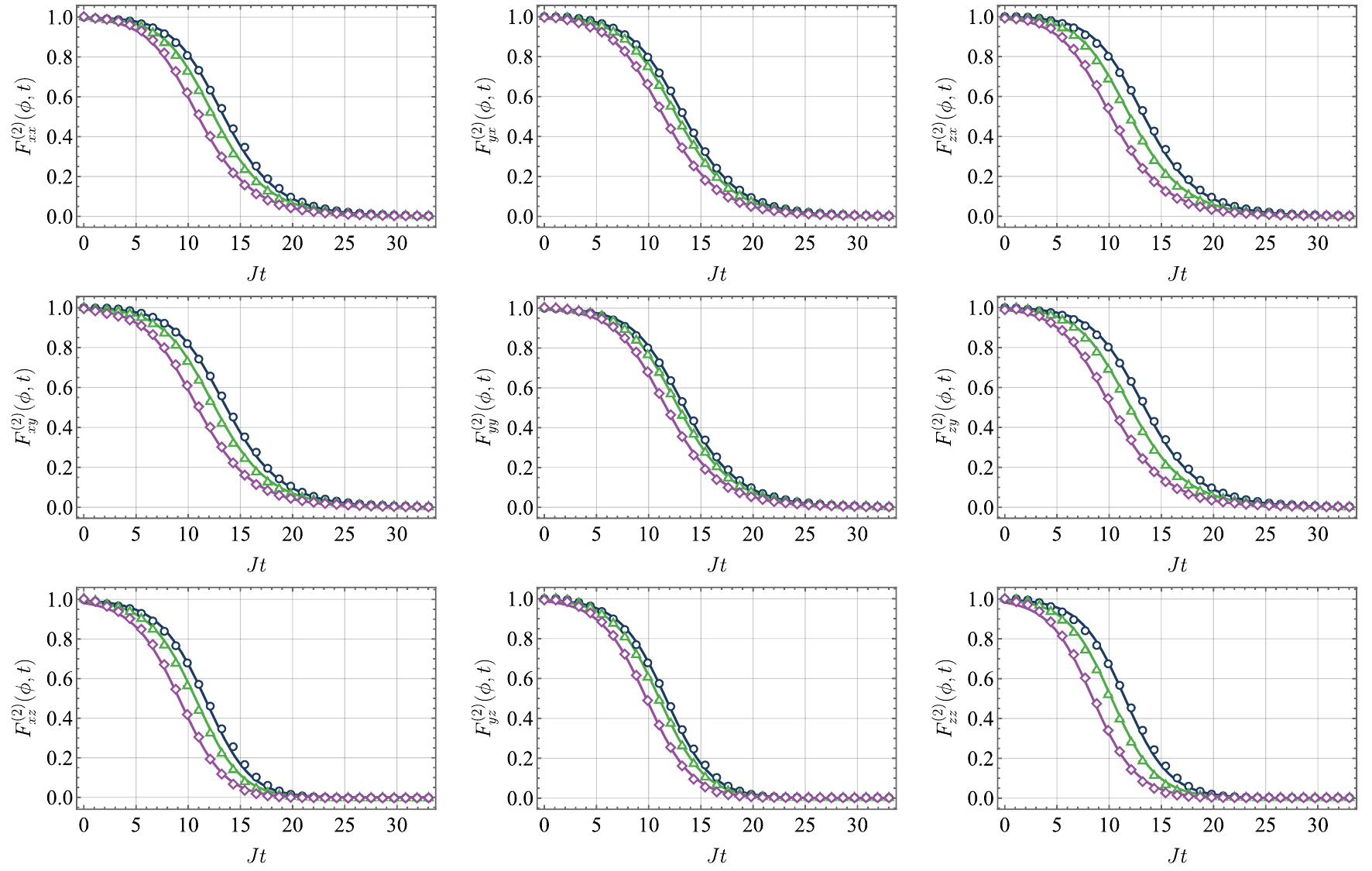}
    \includegraphics[width=0.7\linewidth]{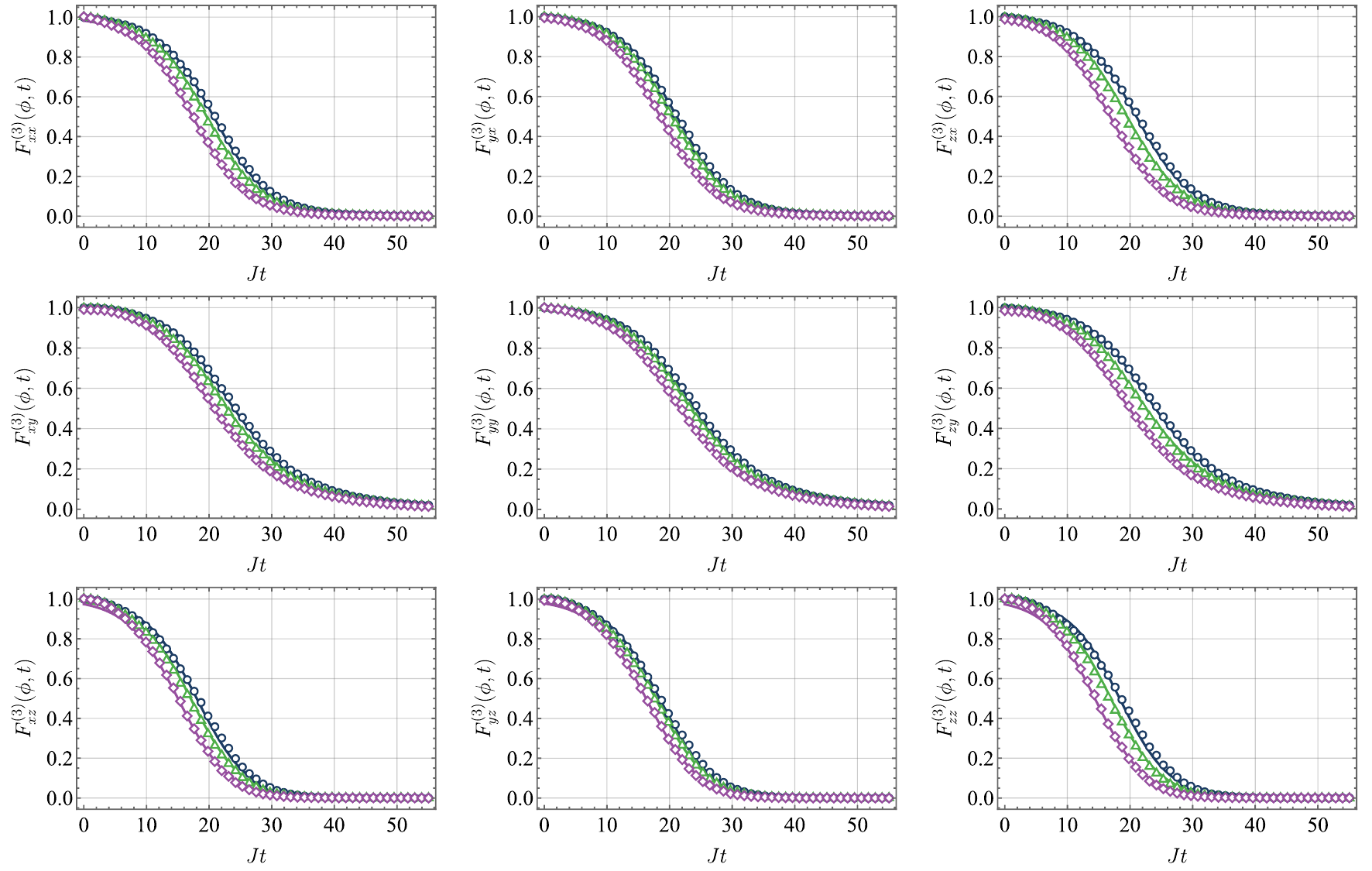}
    \caption{\textbf{The optimal fit for the MQC data using Eq.~\eqref{eqn:main_scramblon}.} We illustrate $F^{(\lambda)}_{\alpha\gamma}(\phi,t)$ for three different $\phi=0$, $\pi/64$ and $\pi/32$. The superscript $\lambda$ labels three different Hamiltonians. The solid curves are fitting results with the scramblon ansatz.}
    \label{fig:optimal_fit}
\end{figure}

\section{Mitigation of the experimental data}

In this section, we present the data obtained using our mitigation method. For illustrative purposes, we use Hamiltonian 1, as defined in the main text. We also provide a detailed analysis of the errors.

\begin{figure}
    \centering
    \includegraphics[width=0.7\linewidth]{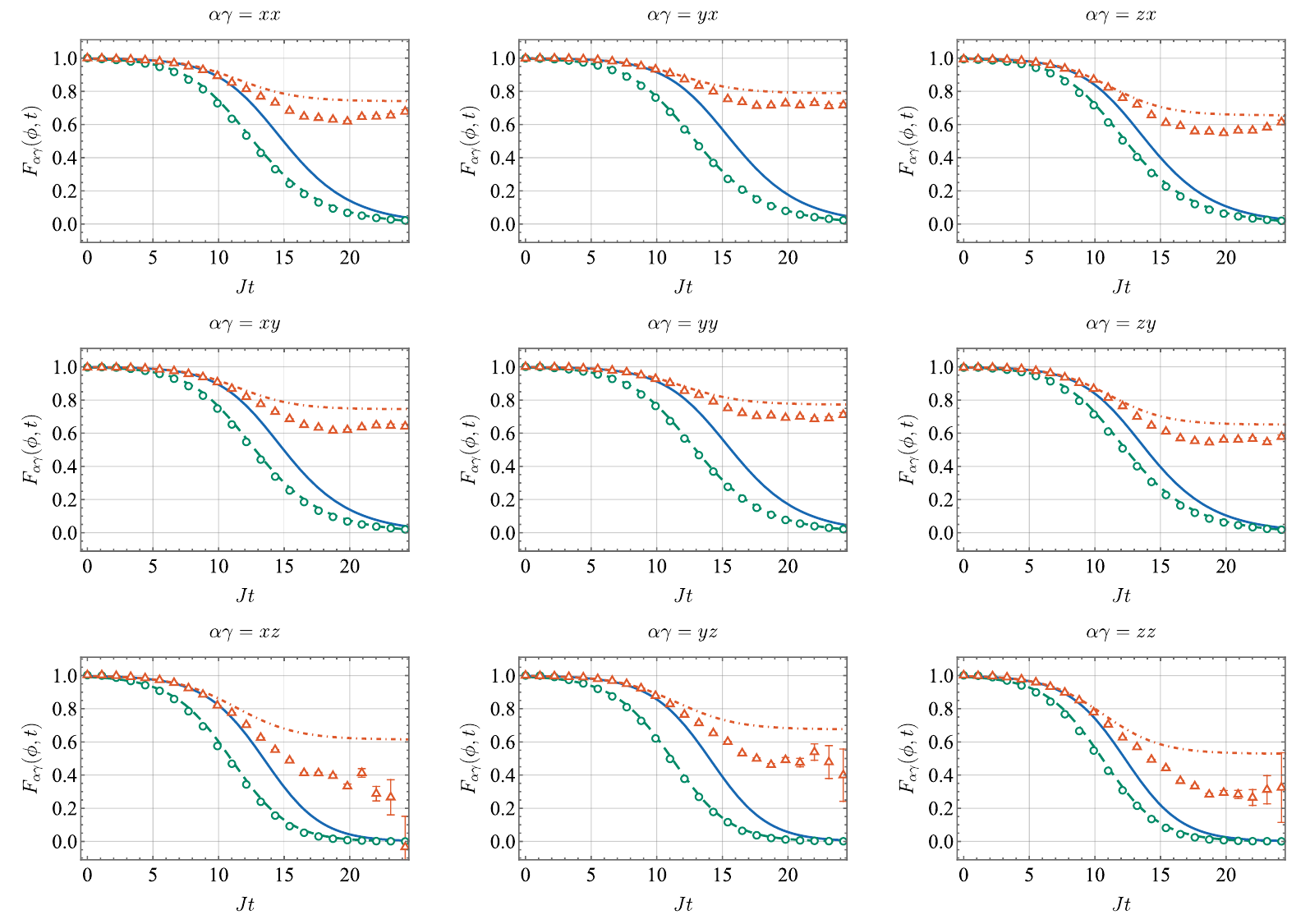}
    \includegraphics[width=0.7\linewidth]{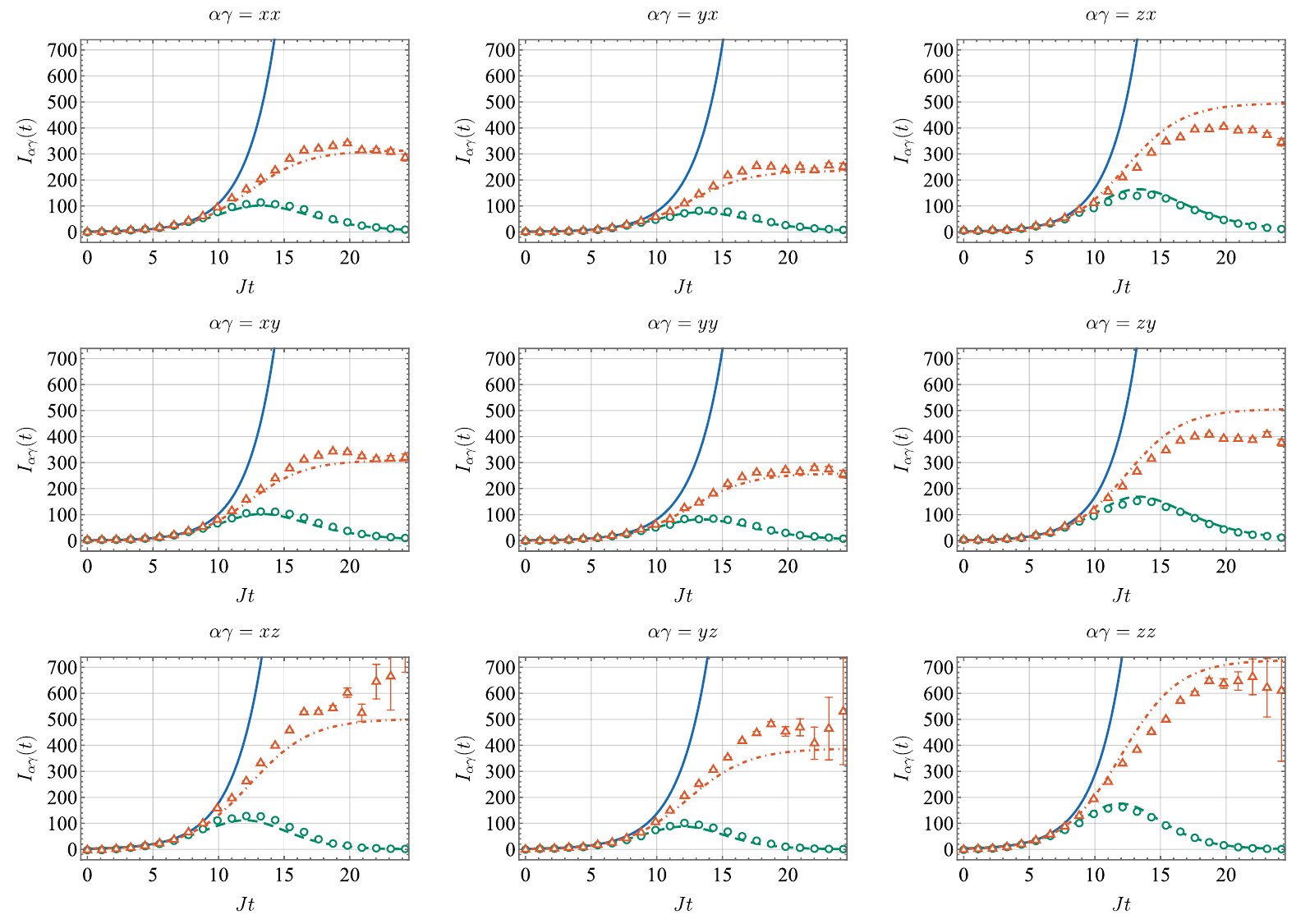}
    \caption{\textbf{Mitigation of MQC Data with Hamiltonian 1.} The system parameters are identical to those in Fig. 4 of the main text, with the exception that more $\alpha\gamma$ directions are considered.}
    \label{fig:mitigation_fit}
\end{figure}

As previously discussed, the standard deviation of the raw data is $10^{-4}$, which is why the error bars for the green circular data points in the main text are not visible. However, when the signal decays to the order of $10^{-4}$, the error bars for the normalized data $\tilde{F}_{\alpha\gamma}$ become significant. This is a direct consequence of error propagation in division, specifically for a ratio of the form $\frac{\mathcal{A}\pm \sigma _\mathcal{A}}{\mathcal{B}\pm \sigma _\mathcal{B}}$. The propagated error for the ratio is given by $\frac{\mathcal{A}}{\mathcal{B}}\pm \sqrt{\frac{\sigma _\mathcal{A}^2}{\mathcal{B}^2} + \frac{\mathcal{A}^2 \sigma _\mathcal{B}^2}{\mathcal{B}^4}}$. When the denominator $\mathcal{B}$, for example $F_{\alpha\gamma}(0,t)$, becomes small, the error bar of the ratio becomes significantly larger. This effect is clearly observed in Figure \ref{fig:mitigation_fit}, where the error bars for the orange triangular data points are notably large.

Finally, we outline the procedure for obtaining the OTO commutator $I_{\alpha\gamma}(t)$ from the raw data of $F_{\alpha\gamma}(\phi,t)$. To ensure numerical stability, we obtain the second derivative by fitting the data with the formula $F_{\alpha\gamma}(\phi,t) \sim d_0 + d_1 \phi^2 + d_2\phi^4 + d_3 \phi^6$. The OTO commutator is then determined from the best-fit parameter as $I_{\alpha\gamma}(t) = -2 d_1$.

Detailed information regarding our error processing methodology can also be found in the accompanying Mathematica script, \texttt{draft\_fig4.nb} \cite{code_zenodo}.

\newpage

\bibliography{ref}
\bibliographystyle{naturemag}